\documentclass[12pt]{article}

\input epsf
\def\beq{\begin{eqnarray}}
\def\eeq{\end{eqnarray}}
\def\m{M_*}
\def\mpl{M_{\rm Pl}}

\def\lsim{\mathrel{\rlap{\lower3pt\hbox{\hskip0pt$\sim$}}
     \raise1pt\hbox{$<$}}}         %less than or approx. symbol
\def\gsim{\mathrel{\rlap{\lower4pt\hbox{\hskip1pt$\sim$}}
     \raise1pt\hbox{$>$}}}         %greater than or approx. symbol

\usepackage{amsmath}
\usepackage{amsfonts}

\begin{document}

\begin{titlepage}

\begin{flushright}
{NYU-TH-05/12/10}
\end{flushright}
\vskip 0.9cm

\centerline{\Large \bf  Perturbations of Self-Accelerated Universe}

\vskip 0.7cm
\centerline{\large C\'edric Deffayet$^{a,b}$, Gregory Gabadadze$^c$ and 
Alberto Iglesias$^c$}
\vskip 0.3cm
\centerline{\it $^a$APC \footnote{UMR 7164 (CNRS, Universit\'e Paris 7, 
CEA, Observatoire de Paris)}, 11 place Marcelin Berthelot,}
 \centerline{\it 75005 Paris Cedex 05, France}
\centerline{\it $^b$ GReCO/IAP \footnote{UMR 7095 (CNRS, Universit\'e 
Paris 6)}, 98 bis Boulevard Arago,}
\centerline{\it 75014 Paris, France }
\centerline{\em $^c$Center for Cosmology and Particle Physics}
\centerline{\em Department of Physics, New York University, New York, 
NY, 10003, USA}

%\vskip 1.9cm

\begin{abstract}

We discuss small perturbations on the 
self-accelerated solution of the DGP model, 
and argue that claims of instability of the solution 
that are based on linearized calculations are 
unwarranted because of the following: 
(1) Small perturbations of an empty self-accelerated 
background can be quantized consistently without yielding 
ghosts. (2) Conformal sources, such as radiation, 
do not give rise to  instabilities either. 
(3) A typical non-conformal source could  introduce ghosts in the
linearized approximation and become unstable, however, it 
also invalidates the  approximation itself. Such a 
source creates a halo of variable curvature that locally dominates 
over the self-accelerated background and extends 
over a domain in which the linearization breaks down. 
Perturbations that are valid outside the halo may not continue 
inside, as it is suggested by some non-perturbative solutions.
(4) In the Euclidean continuation of the theory, with arbitrary sources, 
we derive certain  constraints imposed by the second order equations 
on  first order perturbations, thus restricting  the linearized solutions that 
could be continued into the full nonlinear theory.
Naive linearized solutions fail to satisfy the above constraints. 
(5) Finally, we clarify in detail subtleties associated with  the 
boundary conditions and analytic properties of the Green's functions.

\end{abstract}

\vspace{3cm}

\end{titlepage}

\newpage

\section{Introduction and summary}

The accelerated expansion of the Universe could be 
due to modification of gravity at 
cosmological distance scales. One ``existence proof'' for such a scenario 
is the DGP model \cite {DGP}, in which 
the self-accelerated solution \cite {Cedric} can be used to 
describe the observed speed-up of the cosmic expansion \cite {DDG}.

While there are a number of interesting works 
exploring the consistency  of the self-accelerated 
solution with the data, see, e.g. 
\cite {DDG}--\cite{Carroll}, the linearized  perturbations
on this background were claimed to contain a ghost-like 
mode \cite {Luty,Ratt,Koyama,Gorbunov,KaloperRuth}. 

If the results of the linearized calculations were reliable,
this would exclude the self-accelerated background from 
a set of physically acceptable solutions of the DGP model.
However, we will show below that the issue is more involved.
Small perturbations on an {\it empty} 
self-accelerated background can be quantized without yielding 
ghosts. Moreover, for conformal sources, such as radiation,
the small perturbations are still ghost free. 
Non-conformal sources, on the other hand, 
could introduce ghost-like instabilities in the 
linearized theory.  However, in the latter case  
the results of the linearized  approximation are unwarranted; 
no conclusion on the (in)stability of the solution of the DGP model can be 
drawn without invoking non-linear dynamics. 

An important role of non-linearities in this model have been emphasized 
in a somewhat different context.  In a simplest case 
of gravity of a static source, e.g., the Sun, it is the  
breakdown of naive linearization  that enables one to 
evade \cite {DDGV} the van Dam-Veltman-Zakharov (vDVZ) 
discontinuity \cite {vDVZ}, which, if present, would have ruled out the 
model already by the Solar system data, without any further need to 
consider cosmology. 

Fortunately, the breakdown of the linearized
perturbation  theory does not necessarily mean {\it incalculability}. 
Instead, an expansion in other small parameters 
should be carried out, see, e.g., 
\cite {DDGV},\cite {Gruzinov},\cite {Kaloper}. 
For instance, despite the fact that 
the nearby gravitational field produced by a static source
(such as Sun, Earth, etc.) is incalculable within the conventional 
linearized approximation, the alternative approaches give rise to 
a {\it weak} field which approximates the result of 
General Relativity (GR) to a high degree of accuracy, with 
tiny deviations that could be potentially measurable (see  
\cite {DGZ,Lue} and \cite {GI}). 
The fundamental reason for this is the absence of a  
continuous transition from DGP to GR in the linearized theory (i.e., 
the vDVZ discontinuity is present there), 
whereas, in a full non-linear model such a transition does exist, 
at least in exact solutions of the classical theory \cite {DDGV}.

After these general remarks, we turn below to a brief 
description of the content and  main results of the present work, 
with all the important technical details left  
for the bulk of the paper and Appendices.

We start in Section 2 by studying  an empty self-accelerated solution, 
i.e., the background on which no external matter/radiation
or non-linear gravitational sources are introduced. This is 
an artificial reduction of the theory, but it serves as an instructive 
playground. We look at small gravitational perturbations on this background. 
The linearized approximation gives a well-defined spectrum of 
tensor perturbations, but in addition, there is a scalar mode 
that could or could not give rise to the ghost-like instability. 
We will show that for the empty brane the scalar acts as a Lagrange 
multiplier. Such a theory
is analogous to conventional QED in which the Lorentz gauge-fixing condition 
is introduced by the Lagrange multiplier technique.
It has been known that such a model can be quantized consistently,
e.g., in the canonical approach by the Nakanishi-Lautrup 
method \cite {Nakanishi,Lautrup},  or more straightforwardly, 
by the path integral approach. Likewise, in the gravitational case at hand, 
we will show that the perturbations can be quantized 
in a path integral approach without negative norm states (i.e., without 
the ghost). We will discuss why our results differ from 
those conclusions of Ref. \cite {Gorbunov} that deal with the empty 
self-accelerated solution.

Things change  when  non-conformal  matter  or non-linear 
gravitational sources are  
allowed in the theory.  For simplicity we consider below 
an external matter source $T_{\mu\nu}$. The linearized calculations 
suggest that $T_{\mu\nu}$ with a nonzero trace 
could excite a ghost-like instability\footnote{This 
statement, even within the linearized approach itself,  
needs further clarification; for details see calculations 
and discussions in Section 4.}. If this were all, then
any nonzero $T$  could start to radiate the ghost-like mode, 
rapidly destabilizing itself.

The above logic, however, misses the main feature of the DGP model,
the importance of the nonlinear interactions. 
To put our discussions in 
a context let us first consider an empty brane with 
flat space-time as a background (this is not a self-accelerated 
solution yet!). Moreover, consider for simplicity a spherically
symmetric matter distribution  of radius $r_0$ 
and density $\rho_0$. The 4D gravitational radius of this source is  $r_M = 
2 G_N M = 2 G_N (4/3)\pi r_0^3 \rho_0$.  There is a new 
distance scale in the theory \cite {DDGV} 
(called the Vainshtein scale \footnote{A similar scale was first 
found by Vainshtein in the context of 4D massive gravity \cite {Arkady}})
\beq
r_*\sim \left (r_M/H_0^2   \right )^{1/3},
\label{r*}
\eeq
where $H_0\simeq 10^{-42}$ GeV is the present-day value of the 
Hubble parameter. Below the scale (\ref {r*}), i.e., for $r\lsim r_*$ 
the linearized approximation breaks down \cite {DDGV}. 
For this breakdown to be relevant,
$r_* $ has to be bigger that $r_0$. This 
translates into the condition on the density of the source
$\rho_0\gsim \rho_c$, where $\rho_c \sim 10^{-29} g/cm^3$
is a critical density of the Universe. For such sources the
invariant scalar curvature, $R$, is bigger than the space-time 
curvature of the observable Universe, i.e., $R\gsim H_0^2$. 
Most of the sources around us satisfy this condition; 
we will refer to them as  {\it supercritical} sources. In any 
realistic setup, and on the self-accelerated background 
in particular,  the scalar curvature due to the background is 
already ``critical'' $R\sim H_0^2$; therefore,  within any overdensity on top 
of this background, the local curvature will be  supercritical 
$R \gsim H_0^2$, and one should expect the naive linearization to 
break down. One could certainly think of a 
case when one ``prepares by hand'' a subcritical
source by distorting locally the self-accelerated background. 
Such a source, however, would by itself be unstable.
One way  for this instability to develop, is for the source
to reduce its size until its density becomes supercritical. 
This is energetically favorable since the bare mass of the 
supercritical sources gets screened \cite {GI}:
A supercritical  source responds to the  breakdown  of 
the linearized approximation by creating a halo of a variable curvature 
which extends over a domain or size $r_*$.  
This halo screens the stress-tensor of the source. Moreover, 
the local curvature in the halo  dominates over the curvature 
of the self-accelerated background itself \cite {GI}. 

How can nonlinear-effects 
change the conclusions of the linearized theory?  There seem to be at 
least two ways, summarized in the paragraphs [A] and 
[B] below, and discussed in detail in Section 3.

[A] The non-perturbative halo that extends to $r_*$ renormalizes
the stress-tensor of the source, as seen from distances larger than
$r_*$. Hence, there is a question  of matching of the linearized  modes  
obtained for $r>r_*$ to ``non-perturbative modes'' that could potentially be 
calculated at $r<r_*$. {\it A priori}, there is no reason to 
believe that all the linearized solutions can be continued inside 
the $r_*$  region, and yet this is an assumption made implicitly 
in the works claiming the instability of the self-accelerated 
background.  It may well be that certain 
perturbative solutions of the exterior region cannot be smoothly 
continued inside  $r_*$.  An evidence in favor of this 
is a  solution of \cite {GI}, in which the interior 
non-perturbative metric interpolates smoothly 
into a  non-perturbative solution outside $r_*$, 
but not into the naive linearized solution. Although the gravitational 
fields are weak  in both interior and exterior regions, they have non-analytic 
dependence on couplings/parameters of the theory and cannot be obtained in the 
linearized approximation. 

[B] There is a  known phenomenon of 
``linearization instability'' in theories of gravity \cite {BrillDeser}, see  
also \cite {Mont,Kastor,Higuchi1,Susskind}. It is said to occur  when not all 
the solutions of the linearized approximation can be carried over 
into the full nonlinear theory. Formally speaking, it arises 
because the nonlinear equations impose certain constraints
on the results of the linearized  calculations. These constraints cannot 
be captured in the linearized theory itself, and the linearized solutions 
typically fail to satisfy them. The fundamental reason why these 
constraints arise is this: Gauge fixing leaves the residual 
freedom  w.r.t. the gauge transformations by  the gauge functions that coincide
with the Killing  vectors of the corresponding background.  Then, there 
could exist a charge, that is an integral of a Killing vector contracted 
with a certain current, which is a generator of these transformations. 
Upon quantization of the theory, the physical states should be 
required to be zero eigenvalue eigenstates  of  the charge 
operator (the  charge is zero  in the classical theory by 
construction).  Furthermore, the requirement  that this charge 
be zero order-by-order in the weak field expansion, imposes in the second 
order perturbation  theory an integral constrain  on the purely first 
order results.

The above described phenomenon is rather generic; the only 
condition  it requires is that the charge itself be 
well-defined (e.g., the integral be convergent). This is certainly the  case 
if the background space-time has a compact spatial section, then 
the integral in the definition of the charge is w.r.t. this 
compact 3-space and its convergence is guaranteed. In this case, the physical 
meaning of the additional constraint is simple -- a compact 3-space 
cannot support an  isolated ``gauge charge'', since  its flux lines have 
nowhere to go.

On the self-accelerated background, similar global constraints  
can be found in its Euclidean continuation. 
In this case, the analog of a compact 3-space of the GR 
examples is the $S^{4}$ of the euclideanized worldvolume 
de Sitter (dS)  space. However,  
there is also an additional  nonlinear equation  on 4D quantities 
that exists in DGP but  does not arise in GR. 
Complementing this equation  with the compactness of  $S^{4}$,  one finds a 
severe  global constraint on linearized perturbations,
as we will show in Section 3.

At the moment of this writing, we do not know whether the nonlinear 
effects discussed in paragraphs [A] and [B]  improve or worsen the 
stability.  however, the following is certain: 
Any claims on the (in)stability of the 
self-accelerated  solution are unjustifiable 
without studying non-linearities. 
We emphasize these arguments here partially in response to the    
dissemination of an implicit (and erroneous in our opinion) 
assumption that  the question  of (in)stability in this model 
can be addressed just by looking at the linearized perturbations.

The right way to do calculations is different from what's usually done 
in GR. For a given cosmological source/overdensity, as a first step, 
one could calculate the perturbations in the domain $r\lsim r_*$.  There, 
it is the metric due to the source that dominates over the
cosmological background. These perturbations are the 
most relevant ones, e.g., for the structure formation.  
One could use the  small ``graviton mass expansion'' about the background 
of the source to see the presence of (in)stability of the theory. 
That the leading order results should be close to those of GR is 
implied by the expansion itself, however, this cannot guarantee the 
consistency of the results (i.e., absence of ghost and other instabilities) 
in the sub-leading order. These are the checks that should be done.

The above comments also concern attempts in the literature 
to contrast the predictions of the model with observations.  In some of 
those cases, an assumption is being made that the values of the 
cosmological parameters can be extracted from  the early  cosmology using 
conventional GR, and those results could then 
be  used to constrain the late time acceleration in DGP, see, e.g.,
\cite {Goobar}.  This  however is not well justified: 
The structure formation in DGP, although is expected to be 
qualitatively close to that of  GR \cite {Lue2}, is nevertheless 
influenced by all the non-linear effects described 
above, and  will differ from the GR results when it comes 
to the precision tests (see, e.g., comments in \cite {Maa}).

In the second part of the paper  we discuss the 
DGP model with non-conformal sources 
which is artificially truncated at the linearized level.
The resulting theory is ruled out already by the Solar system data,
and presents only an academic interest; 
yet this is the model implicitly adopted in the works 
which ignore nonlinear effects while discussing the 
question of stability. We show that even in this 
model  there are issues that have not been clarified. 
The main one is that of the boundary conditions and 
poles of the Green's functions from which the content of the physical 
states of the theory can be read off. 
This involves subtleties which  we discuss in Section 4.    

We do not discuss the quantum gravity loops in the 
present paper. The loop corrections lead to the breakdown of the 
perturbation theory, like the tree level diagrams do. 
The scale of the breaking is rather low, and one may worry about the higher 
dimensional operators that can be generated at that low scale and spoil the 
predictivity of the theory \cite {Luty,Ratt}. 
The self-accelerated solution itself is right at that breaking 
scale, and any perturbation about it, exceeds that scale.
If the above  approach is adopted, then the analysis  
showing that on the self-accelerated background the bending mode flips
the sign of its kinetic term, and, hence, produces a ghost 
\cite{Ratt}, cannot be conclusive. These issues, and ways to address them 
consistently, were discussed by Dvali  \cite {Dvali,Dvalitalk}. 
In the present paper we adopt the 
approach \cite {GI} in which all the  loops are calculated  about 
some external classical background field (e.g., early cosmology, 
galaxies, planets etc). In this case, the loop effects are
additionally suppressed by the physical scale of the background metric, 
and can be  harmless at observable distances.

\section{Perturbations of empty self-accelerated brane}

In this Section we discuss an instructive
setup with an {\it empty} self-accelerated brane. 
Gravity is truncated to linearized perturbations about 
the background that is a solution of the full nonlinear equations. 
There are no sources placed on such a brane; the cosmic acceleration
is triggered and maintained by the gravitational sector itself, 
and we look at the evolution of small gravitational 
perturbations on such a background. The obtained results
will also be applicable to the theory with conformal sources. 

\vspace{0.1in}

The DGP gravitational action takes the form  \cite {DGP}
\beq
S\,=\,{\mpl^2 \over 2} \,\int\,d^4x\,\sqrt{g_4}\,R(g_4)\,+
{\m^{3} \over 2}\,\int \,d^4x\,dy\,\sqrt{g}
\,{\cal R}(g)\,+...,
\label{1}
\eeq
where $R$ and ${\cal R}$ are the four-dimensional
and five-dimensional Ricci scalars, respectively, and
$\m$ stands for the gravitational scale
of the bulk theory. The analog of the graviton mass
is $m_c= 2\m^3/\mpl^2$. The $(4+1)$ coordinates are $x^M=(x^\mu,\ y)$, $\mu=0,
\dots, 3$;  $g_4$ and $R$ are the absolute value of the 
determinant and 4D curvature for the 4D 
components $g_{4\mu\nu}(x,y)$ of the 5D metric
$g_{AB}(x,y)$
\beq
{g}_{\mu\nu}(x,y=0)\equiv g_{4\mu\nu}(x)\,.
\label{bargg}
\eeq
There is a boundary (a brane) at $y=0$ and ${\bf Z}_2$
symmetry across the boundary is imposed. The 
boundary Gibbons-Hawking term is implied to warrant  
the correct Einstein equations in the bulk.
The matter fields, that are omitted in the present section, 
will later be assumed to be localized on a brane.
 
Below we consider  the metric perturbations of the form:
\beq
ds^2 = \left ( {\bar g}_{\mu\nu}(x,y) + \delta g_{\mu\nu}(x,y) \right )
dx^\mu dx^\nu + 2 \delta g_{\mu 5}dx^\mu dy+ 
\left (1+ \delta g_{55}(x,y)   \right )dy^2\,, 
\label{metricg}
\eeq
where the background quantities will be denoted by the bar
$ {\bar g}_{\mu\nu}(x,y) =A^2(y) \gamma_{\mu\nu}(x)$.
We start with the 5D equations of motion. They decompose into 
$\{\mu 5\}$, $\{5 5\}$ and $\{\mu\nu \}$ components.
Let us first look at the $\{\mu 5\}$, $\{5 5\}$ equations.
They read respectively 
\beq
\nabla^\mu K_{\mu\nu} = \nabla_\nu K\,,
\label{mu5new} \\
R= K^2 -K_{\mu\nu}K^{\mu\nu}\,.
\label{55} 
\eeq
Here, $K=g^{\mu \nu}K_{\mu \nu}$ is the trace of the 
extrinsic curvature tensor defined as follows: 
\beq
K_{\mu \nu}\,=\,{1\over 2N}\,\left (\partial_y g_{\mu \nu} -\nabla_\mu 
N_\nu
-\nabla_\nu N_\mu \right )\,,
\label{K}
\eeq
and $\nabla_\mu $ is a covariant derivative w.r.t. the metric $g_{\mu \nu}$.
We introduced the {\it lapse} scalar field  $N$, and 
the {\it shift} vector field $N_\mu $ according to the standard 
rules:
\beq
g_{\mu 5}\,\equiv \, N_\mu =g_{\mu\nu}N^\nu\,,~~~g_{55}\,\equiv \,N^2\,+\,
g_{\mu\nu}\,N^\mu \,N^\nu\,. 
\label{adm}
\eeq 
Equations (\ref {mu5new}) and (\ref {55}) 
should be satisfied everywhere in the bulk as well as  
on the brane. The $\{\mu\nu \}$ equation, on the other hand,
takes different form depending whether it is in the bulk or on the brane
\beq
&G^{(5)}_{\mu\nu}& =0\, ~~~{\rm for}~~~y\neq 0\,, \label{bulkmunu} \\
&G_{\mu\nu}|_{y=0}& - {m_c\over 2} 
[ K_{\mu\nu} - g_{\mu\nu} K]^{+\epsilon}_{-\epsilon} = T_{\mu\nu}(x)\,.
\label{junction}
\eeq

\subsection*{Spectrum of empty self-accelerated brane}

We begin by  analyzing small perturbations about an {\it empty} (i.e., 
$T_{\mu\nu} =0$)  self-accelerated brane.  
We introduce the following self-explanatory notations
\beq
g_{\mu\nu} = {\bar g}_{\mu\nu} +\delta g_{\mu\nu}\,,~~~K_{\mu\nu}
= {\bar K}_{\mu\nu} + \delta K_{\mu\nu} \,, \\
{\bar K}_{\mu\nu} =n {\bar g}_{\mu\nu} \,,~~~ {\bar K} = 
{\bar g}^{\mu\nu} {\bar K}_{\mu\nu} =4 n\,~~~n\equiv {\partial_y A\over A}\,,
\label{fluct}
\eeq
where, according to the above definitions: 
\beq
K = g^{\mu\nu}K_{\mu\nu} \equiv {\bar K} + \delta K\,, 
~~~ \delta K = {\bar g}^{\mu\nu}\delta K_{\mu\nu}
-  n {\bar g}^{\mu\nu} \delta g_{\mu\nu}\,. 
\label{Knew}
\eeq 
For the self-accelerated solution $|H|=m_c$.
The unperturbed part of the off-diagonal equation (\ref {mu5new}) 
is identically satisfied. While the unperturbed parts of the 
$\{55\}$ and junction equations respectively give: ${\bar R}= 12H^2$
and ${\bar R}= 12H m_c$; these are consistent with each other
(from now on we will replace $|H|$ by $H$ with the understanding
that we should always take a positive value for the latter
no matter what is the direction of the time arrow).

Let us analyze the small fluctuations 
described by the above system of equations. 
The perturbations of the off-diagonal equation (\ref {mu5new}) 
satisfy  the following relation
\beq
\nabla^\mu \left (  \delta K_{\mu\nu} - n \delta g_{\mu\nu}\right )=
 \nabla_\nu \left ({\bar g}^{\rho\sigma} \delta K_{\rho\sigma} -
n {\bar g}^{\rho\sigma} \delta g_{\rho\sigma} \right )\,,
\label{mu5pert}
\eeq
where $\nabla^\mu$ is a 4D covariant derivative
constructed out of ${\bar g}_{\mu\nu}$. We will return to this 
equations shortly.

For small perturbations of the $\{55\}$ equation (\ref {55}) we find
\beq
\delta R = 6 n ( {\bar g}^{\mu\nu}\delta K_{\mu\nu}
-n {\bar g}^{\mu\nu} \delta g_{\mu\nu})\,,
\label{55pert}
\eeq
while from the source-free junction conditions one gets
\beq
\delta R|_{0^+} = 3 m_c ( {\bar g}^{\mu\nu}\delta K_{\mu\nu}
-n {\bar g}^{\mu\nu} \delta g_{\mu\nu})|_{0+} \,. 
\label{junctpert}
\eeq
Since $m_c=H$, equations (\ref {55pert}) and (\ref {junctpert})
are in contradiction unless the r.h.s. of (\ref {junctpert}) is zero.
Moreover, requiring the continuity of the 4D curvature perturbations 
$\delta R$  we  conclude that the r.h.s. of  (\ref {55pert})
should also be zero for arbitrary $y$:
\beq
{\bar g}^{\mu\nu}\delta K_{\mu\nu} -n  {\bar g}^{\mu\nu} 
\delta g_{\mu\nu}=0\,. 
\label{sourcefree}
\eeq
We will come back to this relation below. 

Furthermore, we need to study small perturbations of the bulk $\{\mu\nu \}$ 
equation. Since we are dealing with an empty brane (or the one with 
a conformal stress tensor), we can adopt
the Gaussian  normal coordinates and choose the 
following ``gauge''\footnote{In general, this is not an acceptable 
gauge fixing condition  for the metric that couples to a non-conformal 
source. That is why one eventually needs to introduce the brane 
bending mode \cite {GT}.}
\beq
\nabla^\mu \delta g_{\mu\nu} =0\,,~~~ {\bar g}^{\mu\nu}\delta 
g_{\mu\nu} =0\,,~~~ \delta g_{55}= \delta g_{\mu5} =0\,.
\label{ttgauge}
\eeq
In this gauge the bulk $\{\mu\nu \}$ equation takes 
the form:
\beq
\partial_y^2 \delta g_{\mu\nu} + {1\over A^2} 
\left (\square_4 -4H^2 \right )\delta g_{\mu\nu} =0\,.
\label{bulkpert}
\eeq
The junction condition reads
\beq
-{1\over 2} \left (\square_4 -4H^2 +H\partial_y \right )\delta
g_{\mu\nu}|_{0^+}=0\,.
\label{junction00}
\eeq
Let us discuss the KK modes. We use the following decomposition:
\beq
\delta g_{\mu\nu} (x,y) 
\equiv \int h_{\mu\nu}^{(m)}(x) {\tilde f}_m(y) 
dm\,,~~~~~( \square_4 -2H^2 )h_{\mu\nu}^{(m)}(x)  =m^2 
h_{\mu\nu}^{(m)}(x)\,.
\label{KK}
\eeq
It is useful to turn to a 
new coordinate $z$ and a  function $u_m$
\beq
dz \equiv {dy \over A(y)}\,, ~~~~ {\tilde f}_m \equiv A^{1/2} u_m \,. 
\label{new}
\eeq
This choice is very convenient since the bulk equation (\ref {bulkpert})
and the junction conditions (\ref {junction00}) for these variables
reduce respectively to Eqs. (\ref {sbulku}) and (\ref {sjunctionu})
of Appendix A, where the latter equations have been solved
(see also \cite {Koyama}). 
The spectrum consists of a normalizable localized mode
of mass $m_*^2 = 3m_cH-m_c^2 =2m_c^2$ and a continuum of 
KK modes with masses $m_{KK}^2\ge 9H^2/4$. 
Note that because of the boundary condition that depends on the 
KK mass, the modes themselves
are not orthogonal. The physical modes are their linear superpositions. 
However, this won't be important for our discussions.

Besides these states, there is a zero-mode
solution,  $m=0$, ${\tilde f}_0=A^2 $ (see Appendix A for details), 
supported by the above system of equations. This solution is not 
plane-wave normalizable. Its wave-function grows in 
the bulk as fast as the background metric itself.
When  the warp-factor is scaled out, 
the zero-mode has a constant  profile in the $y$ direction,
$A^2(y)\gamma_{\mu\nu}(x) + A^2(y)h^{\rm zm}_{\mu\nu}(x)$,
where $\gamma_{\mu\nu}(x)$ denotes the 4D dS metric and 
$h^{\rm zm}_{\mu\nu}(x)$ is the zero-mode wave-function.  
We will discuss properties of  this mode in Sections 4, 
but we ignore it for the time being.

Let us concentrate on the lowest massive localized mode mentioned above.
The mass of this mode is related to the 
background space-time curvature as follows:
\beq
m_*^2 ={{\bar R}\over 6} = 2H^2\,.
\label{mR} 
\eeq
It has been known for some time that 4D massive gravity on a dS
background with the relation (\ref {mR}) exhibits an additional  
local symmetry \cite {Deser}
\beq
\delta h^*_{\mu\nu}=(\nabla_\mu \nabla_\nu +H^2 \gamma_{\mu\nu})\alpha(x),
\label{sym*}
\eeq
with an arbitrary $\alpha$. This   
enables one to gauge-remove one out of the five on-shell polarizations 
of the massive graviton. The resulting graviton has four physical 
polarizations \cite {Deser}. Advantages and disadvantages of this theory 
are summarized in Appendix B. For further convenience, we will denote 
the Lagrangian of that theory  by ${\cal L}_*$. The question that we'd 
like to discuss is  whether the DGP Lagrangian  on the self-accelerated 
solution  reduces at low energies to ${\cal L}_*$ \cite {Koyama,Gorbunov}.   
To answer this question we note that 
the local symmetry (\ref {sym*}) is present in DGP
only for the restricted gauge functions. 
Indeed, both bulk and brane equations of motion   
are invariant, in the absence of the sources 
(or with conformal sources only),  w.r.t. 
the following symmetry transformations 
\beq\label{sym}
\delta g_{\mu\nu}\to \delta g_{\mu\nu}+(\nabla_\mu\nabla_\nu+
H^2\gamma_{\mu\nu})\sigma~,
\eeq
where the field $\sigma$ satisfies
\beq\label{sigma}
(\Box +4 H^2)~\sigma=0~.
\eeq
One can see this by performing straightforward variations of the 
equations presented above  and using the following identities:
\beq
(\Box-4H^2)(\nabla_\mu\nabla_\nu+H^2\gamma_{\mu\nu})({\rm scalar})&=&
(\nabla_\mu\nabla_\nu-H^2\gamma_{\mu\nu})(\Box+4H^2)({\rm scalar})~,\nonumber\\
\nabla^\mu(\nabla_\mu\nabla_\nu+H^2\gamma_{\mu\nu})~({\rm scalar})  &=&
\nabla_\nu(\Box+4H^2)~({\rm scalar})~.
\eeq
In the absence of sources, the symmetry  (\ref {sym}) can be used to 
remove on-shell a potentially dangerous degree of freedom, and fluctuations 
can be quantized consistently without any ghost or other pathologies, 
as we'll see in the next subsection. 

Before we turn to these issues, however, we'd like to clarify 
the following point.  The difference between the low energy DGP and 
the model ${\cal L}_*$ is not entirely clear from our discussions above.
Indeed, the profile of the localized mode $u_{m_*}$ is known, 
it's  4D effective Lagrangian  can be calculated
by ignoring other modes and integrating  the 5D action 
w.r.t. the $y$ coordinate. Doing so, it would naively seem, that  
the low-energy effective Lagrangian on the self-accelerated branch 
coincides with ${\cal L}_*$. However, as was shown in Ref. \cite {Gorbunov},
the model, in general, {\it does not} reduce to 
the ${\cal L}_*$ theory. This is because of the 
following: In the calculation of the present section the brane 
bending mode was ignored, it was simply put equal to zero.
This is consistent for an empty brane (more precisely, as long as $T=0$). 
As a result of this, a certain constraint, which is present in the full 
linearized theory, is not captured. To find this constraint, and to see whether
it distinguishes DGP from ${\cal L}_*$, one should 
work with the bending mode manifestly present in the action, 
as was done in \cite {Gorbunov}. However, as we will show in the 
next subsection, the bending mode appears as a Lagrange multiplier.
As a result, in the absence of sources, both the low-energy 
Lagrangian of DGP and  ${\cal L}_*$ can be consistently 
quantized to contain four on-shell physical polarizations 
without ghosts or tachyons.

\subsection*{Effective Lagrangian approach}

Let us start with an illustrative model discussed
in \cite {Gorbunov}\footnote{The  authors of \cite {Gorbunov}
attribute this model to R. Rattazzi.}. The Lagrangian contains 
one gauge field $A_\mu$ and two scalars, $\varphi_1$ and $\varphi_2$:
\beq
{\cal L} = -{1\over 4} F_{\mu\nu}^2 -{f_1^2\over 2} (\partial_\mu\varphi_1
+ e A_\mu)^2 + {f_2^2\over 2} (\partial_\mu\varphi_2
+ e A_\mu)^2\,,
\label{qedL}
\eeq
where $f_1^2,f_2^2$ are two positive constants. The above Lagrangian
is invariant under the local symmetry transformations
$\delta \varphi_1 =  \delta \varphi_2
=\alpha,~~\delta A_\mu = -\partial_\mu\alpha/e$. These
can be used to gauge-remove  one of the two scalar fields, e.g.,
one can set  $\varphi_2=0$. Following \cite {Gorbunov},
consider the case
\beq
f_1^2 =f_2^2\,,
\label{f1f2}
\eeq
which  presumably mimics physics of the self-accelerated
solution discussed in the present paper\footnote{The case
$f_1^2 \neq f_2^2$ is supposed to 
mimics the physics of a brane on which additional 
tension is introduced.  This is more subtle and will not be
considered in the present work.}. For this particular choice 
the Lagrangian reduces to \cite {Gorbunov}:
\beq
{\cal L} = -{1\over 4} F_{\mu\nu}^2 -{f_1^2\over 2} (\partial_\mu\varphi_1)^2
-f_1^2 e A^\mu \partial_\mu\varphi_1 \,.
\label{ghostL}
\eeq
By calculating  the Hamiltonian of the
helicity-0 sector of the above theory (\ref {ghostL}), 
this model was claimed to have a ghost \cite {Gorbunov}.
        
To look more closely into this issue, let us perform in (\ref {ghostL}) 
the field {\it redefinition}
\beq
B_{\mu} \equiv A_{\mu} + {1\over 2e} \partial_\mu \varphi_1,~~~
\lambda \equiv f_1^2 e \varphi_1\,.
\label{redef}
\eeq
In terms of the new fields the Lagrangian  (\ref {ghostL}) 
takes  the form (up to a  total derivative)
\beq
{\cal L} = -{1\over 4} F_{\mu\nu}^2 + \lambda \, \partial^\mu B_\mu\,,
\label{NL}
\eeq
where $F_{\mu\nu}$ is a field-strength for the $B_\mu$ field now.
This  is a Lagrangian of
a gauge field $B_\mu$  for  which the  Lorentz gauge-fixing
condition $\partial^\mu B_\mu=0$ has been introduced by means of
the Lagrange multiplier $\lambda$.  It is well known 
how to quantize this Lagrangian consistently without the 
appearance of ghosts or other pathologies. For completeness
we present here the summary of the path integral quantization, as well as 
the canonical quantization due to 
Nakanishi and Lautrup (NL) \cite {Nakanishi,Lautrup}.

In the path integral approach we find
\beq
\int [dB][d\lambda] e^{i\int d^4x {\cal L}} = \int [dB]
\delta (\partial^\mu B_\mu\,) 
 e^{i\int d^4x (-{1\over 4} F_{\mu\nu}^2 )}.
\label{path1}
\eeq
The r.h.s. of the above equation is nothing but the 
path integral for QED with the Lorentz gauge-fixing condition
where the Faddeev-Popov determinant $det |\square|$, which is 
trivial in this case, is dropped. Needless to say that this 
path integral does not propagate any negative norm states (ghosts)
and can be straightforwardly completed to a BRST invariant theory
(the Faddeev-Popov ghost in this case are decoupled).

Counting of the on-shell degrees of freedom in the Lagrangian formalism 
is also straightforward; (\ref {NL}) imposes the Lorentz condition
on the gauge field, removing one degree of freedom, yet 
allows for residual gauge transformations 
$\delta B_\mu =\partial_\mu\alpha$, where $\square \alpha=0$.
This can be used to constrain on-shell the second degree of freedom,
rendering two out of the four modes of $B_\mu$.

In the Lagrangian quantization of (\ref {NL})
the subsidiary Nakanishi-Lautrup condition should be 
imposed on physical states 
of the theory: $\lambda^{(-)}|{\rm Phys}>=0$
(here $\lambda^{(-)}$ denotes the negative frequency part of the
field $\lambda$)\footnote{The NL subsidiary condition is an analog of the
Gupta-Bleuler condition in the conventional Lagrangian quantization
of electrodynamics.}.  The theory defined this way has no 
perturbative classical or quantum ghost-like instabilities 
\cite {Nakanishi,Lautrup}\footnote{We point out that the Lagrangian 
(\ref {NL}),  at the classical level,
could have solutions that are not present in conventional electrodynamics
\cite {GabadadzeShang}. However, these solutions form different 
super-selection sectors as they imply the
presence of some additional physics (e.g., sources at infinity, different
boundary conditions, etc.). We won't consider such backgrounds here
since our goal is to study small perturbations on a given background
(in particular, on the self-accelerated background in the gravity case
discussed below).}.  Moreover, although the canonical Hamiltonian
of the theory does contain negative terms, the physical 
Hamiltonian  that acts on states of physical Hilbert space, 
which obey all the constraints and  NL 
subsidiary condition, is in fact 
positive semi-definite.

%\vspace{0.1in}

Let us now turn to the effective Lagrangian approach to the
empty self-accelerated  brane.  The four dimensional effective action for the 
localized and brane bending modes was derived in \cite{Gorbunov}, 
which we follow below. The effective Lagrangian  takes the form
\cite{Gorbunov}:
\beq\label{eff}
{\cal L}_{\rm eff}={\cal L}_*(A_{\mu\nu})-HA^{\mu\nu}(\nabla_\mu\nabla_\nu
-\gamma_{\mu\nu}\Box-3H^2\gamma_{\mu\nu})\varphi-{9H^2\over 4}
\varphi(\Box+4H^2)\varphi,
\eeq
where $A_{\mu\nu}\equiv\delta g_{\mu\nu}(x,y=0)$ denotes the tensor mode 
and  $\varphi$ is the bending mode. Furthermore,   
${\cal L}_*(A_{\mu\nu}) $ denotes the Lagrangian of a 
linearized massive gravity on the dS 
background when the graviton mass and the cosmological constant satisfy 
a special relation  $m_*^2=2H^2$. The explicit  
expression for ${\cal L}_*(A_{\mu\nu})$ is given in 
\cite {Higuchi} (see also Appendix B), but what is important here 
is that ${\cal L}_*(A_{\mu\nu})$ is invariant w.r.t. the 
local  transformation, $ \delta A_{\mu\nu} = 
(\nabla_\mu\nabla_\nu+H^2\gamma_{\mu\nu})\alpha (x)$, for 
an arbitrary $\alpha(x)$. As a result of this symmetry, 
${\cal L}_*(A_{\mu\nu})$ does not retain any physical 
scalar degrees of freedom 
-- the only on-shell modes described by ${\cal L}_*(A_{\mu\nu})$ are 
2 polarizations of a transverse-traceless (TT) tensor and two 
polarizations of a transverse vector \cite {Deser}\footnote{Notice that
for $m_*^2=2H^2$ there exists a unitary massive spin-two representation of 
the dS group with 4 degrees of freedom.}. However, the last two 
terms of ${\cal L}_{\rm eff}$ are not invariant under the above 
local transformation. Hence, the question arises whether this 
reintroduces back the scalar  polarization that could be a ghost.

We will show  below that the results are very similar to 
those of a  simple gauge-field example \cite {Gorbunov} 
considered at the beginning of this 
subsection. In particular, the non-invariant terms in ${\cal L}_{\rm eff}$
can be recast as a product of a Lagrange multiplier with a 
gauge-fixing condition. Just as in the toy example,  
small perturbations of the  resulting theory can 
consistently be quantized, without any ghosts. 
To see this, we perform a redefinition of  
the  $A_{\mu\nu}$ field
\beq
A_{\mu\nu}=B_{\mu\nu}+(\nabla_\mu\nabla_\nu+H^2\gamma_{\mu\nu})~\omega~.
\eeq
This change of variables does not affect ${\cal L}_*$ since 
${\cal L}_*(A_{\mu\nu})= {\cal L}_*(B_{\mu\nu})$. However, the last
two terms of ${\cal L}_{\rm eff}$ in (\ref {eff}) 
do get modified. To find the result of 
this let us now introduce the traceless operator 
$P_{\mu\nu}\equiv\nabla_\mu\nabla_\nu-{1\over 4}\gamma_{\mu\nu}\Box$, that 
satisfies 
$P^{\mu\nu}P_{\mu\nu}({\rm scalar})={3\over 4}\Box(\Box+4H^2)({\rm scalar})$, 
then,
\beq
{\cal L}_{\rm eff}&=& {\cal L}_*(B_{\mu\nu})-HB^{\mu\nu}(\nabla_\mu\nabla_\nu
-\gamma_{\mu\nu}\Box-3H^2\gamma_{\mu\nu})\varphi-{9H^2\over 4}
\varphi(\Box+4H^2)\varphi\nonumber\\
&&-H\omega(\nabla_\mu\nabla_\nu+H^2\gamma_{\mu\nu})\left(P_{\mu\nu}-{3\over 4}
\gamma_{\mu\nu}(\Box+4H^2)\right)\varphi\nonumber\\
&=& {\cal L}_*(B_{\mu\nu})-HB^{\mu\nu}(\nabla_\mu\nabla_\nu
-\gamma_{\mu\nu}\Box-3H^2\gamma_{\mu\nu})\varphi-{9H^2\over 4}
\varphi(\Box+4H^2)\varphi\nonumber\\
&&+H\omega3H^2(\Box+4H^2)\varphi,
\eeq
and setting $\omega={3\over 4H}\varphi$, the effective Lagrangian becomes
\beq
{\cal L}_{\rm eff}&=& {\cal L}_*(B_{\mu\nu})-\varphi H
(\nabla_\mu\nabla_\nu-\gamma_{\mu\nu}\Box-3H^2\gamma_{\mu\nu})B^{\mu\nu}\,.
\label{Leff2}
\eeq
Remarkably, the bending mode 
$\varphi$ acts as a Lagrange multiplier imposing a certain 
constraint\footnote{Notice that $ (\nabla_\mu\nabla_\nu
-\gamma_{\mu\nu}\Box-3H^2\gamma_{\mu\nu})B^{\mu\nu}\sim \delta R$.}.
This is very similar in spirit to the NL form of the 
electrodynamics Lagrangian (\ref {NL}). Hence, our consideration below 
will be similar too. We start with the path integral quantization of this
theory
\beq
\int [dB][d\varphi] e^{i \int d^4x {\cal L}_{\rm eff}} =
\int [dB]\delta ((\nabla_\mu\nabla_\nu-\gamma_{\mu\nu}\Box-
3H^2\gamma_{\mu\nu})B^{\mu\nu}) e^{i \int d^4x {\cal L}_{*}}\,, 
\label{pathq}
\eeq
the r.h.s. of this can be regarded as the path integral of a theory
described by ${\cal L}_{*}$ in which the gauge fixing condition
\beq
(\nabla_\mu\nabla_\nu
-\gamma_{\mu\nu}\Box-3H^2\gamma_{\mu\nu})B^{\mu\nu}=0,
\label{dR0} 
\eeq 
is imposed. Moreover, it is straightforward to calculate that 
the Faddeev-Popov determinant for the gauge condition
(\ref {dR0}) takes a 
form, $det |\square +4H^2|$. This is a trivial 
factor (independent of $B$) which can be restored 
in the r.h.s. of  (\ref {pathq}) at the expense of an arbitrary 
overall prefactor. The result reads
\beq
\int [dB]\delta ((\nabla_\mu\nabla_\nu-\gamma_{\mu\nu}\Box-
3H^2\gamma_{\mu\nu})B^{\mu\nu}) det |\square +4H^2|  
e^{i \int d^4x {\cal L}_{*}}\,. 
\label{pathq1}
\eeq
This is nothing but the path integral for the theory ${\cal L}_*$
with the gauge fixing and Faddeev-Popov determinant terms.
Therefore, we conclude that the quantization of ${\cal L}_{\rm eff}$ 
is identical to the quantization of ${\cal L}_*$ and both of these theories
(in the absence of sources) propagate 4 physical on-shell 
polarizations and no ghosts.  

Let us also count degrees of freedom in the Lagrangian formalism. 
We first recall how things work in the theory described by  
${\cal L}_* $.  The equations of motion 
and Bianchi identities lead to the condition 
$\nabla^\mu B_{\mu\nu} = \nabla_\nu B$ (see Appendix B). This equation, 
as well as the Lagrangian ${\cal L}_* $ itself, 
are invariant under the local transformations 
$ \delta B_{\mu\nu} = (\nabla_\mu\nabla_\nu+H^2\gamma_{\mu\nu})
\alpha (x)$, with an arbitrary $\alpha$. This can be used to 
gauge remove the trace, i.e., set  $B^\mu_\mu=0$. As a result, we also 
get $\nabla^\mu B_{\mu\nu} = 0$. Hence, the $B_{\mu\nu}$ field is 
transverse-traceless, hence, it has only 5 independent components. 
However, there still remains gauge symmetry  under 
the transformations $ \delta B_{\mu\nu} = 
(\nabla_\mu\nabla_\nu+H^2\gamma_{\mu\nu})~
\sigma (x)$, where $(\square +4H^2)\sigma=0$. This allows to 
eliminate one more on-shell degree of freedom. Hence, $B_{\mu\nu}$ field in 
the  Lagrangian  ${\cal L}_* $ describes  only 4  
physical polarizations \cite {Deser}
\footnote{The Lagrangian ${\cal L}_*$, however, {\it does not} permit the 
$B_{\mu\nu}$ field to be coupled to a non-conformal source, 
$T\neq 0$ (see e.g., Appendix B). This is very similar to how  
the conventional electrodynamics does not permit the photon 
to be coupled to a non-conserved current.}.

Let us now see how the counting works for 
${\cal L}_{\rm eff}$. Variation w.r.t. $\varphi$ 
yields (\ref {dR0}). Furthermore, it is straightforward to check that 
the Bianchi identities for the $\{\mu\nu\}$ 
equation lead to $\nabla^\mu B_{\mu\nu} = \nabla_\nu B$.
The divergence of the latter  
combined with (\ref {dR0}) leads to $B=0$,
and, as a consequence, $\nabla^\mu B_{\mu\nu}=0$ too.
Moreover, the Lagrangian ${\cal L}_{\rm eff}$, 
as well as all the above  constraints, 
are still invariant under the transformations
$ \delta B_{\mu\nu} = (\nabla_\mu\nabla_\nu+H^2\gamma_{\mu\nu})~
\sigma (x)$, where $(\square +4H^2)\sigma=0$. These can be used
to eliminate one more degree of freedom of the $B_{\mu\nu}$ field, 
reducing the total number of  its physical on-shell states down to 4.
Therefore, as far as the $B_{\mu\nu}$ field is concerned,
there is no difference between  the description in terms of 
${\cal L}_* $ or  ${\cal L}_{\rm eff}$.
Let us now look at the  $\varphi$  field. 
This field is a Lagrange multiplier in (\ref {Leff2}). 
Moreover, the trace of the  $\{\mu\nu\}$ equation plus the 
Bianchi identities lead  to $(\square +4H^2)\varphi=0$. This is reminiscent of
the condition $\square \lambda=0$ on the Lagrange multiplier 
emerging in (\ref {NL}). As in that case, the value of the Lagrange multiplier
$\varphi$ can be set to zero and this condition should be possible 
to maintained in the full quantum theory by imposing the  NL like 
constraint  on the physical  states  $\varphi^{(-)}|{\rm Phys}\rangle =0$.
This must be sufficient to guarantee that the 
small perturbations described by the Lagrangian 
${\cal L}_{\rm eff}(B)$  yield no ghosts, as it was shown in the 
path integral quantization. 
Moreover, the physical degrees of freedom are identical 
to those described by ${\cal L}_* $.

Summarizing, we showed that  small perturbations
on the empty self-accelerated brane can be quantized consistently 
and do not exhibit any  ghost instability. These results also 
apply to the brane endowed with conformal sources.
This is because all the arguments and counting of the 
degrees of freedom go through in the presence of 
conformal sources, since the symmetry (\ref {sym}),(\ref {sigma}) 
is preserved in that case\footnote{In this paper we 
do not discuss stability of the background w.r.t. possible classical,
and semi-classical non-perturbative effects, such as the decay
of the vacuum etc. If these instabilities are present, 
we'd expect them to set in very slowly, without affecting 
the observations; moreover, the full nonlinear theory should be 
used to study their consequences.}.

The above presented counting of the degrees of freedom in 
${\cal L}_{\rm eff}$ changes once sources 
with  $T\neq 0$ are introduced. Details are given 
in  Section  4; here we just note that although 
the relation $\nabla^\mu B_{\mu\nu} = \nabla_\nu B$ remains intact,
one looses the symmetry (\ref {sym}),(\ref {sigma}) that was 
essential to remove one on-shell degree of freedom. 
The physical metric that couples to the stress-tensor is
\beq
\delta {\tilde g}_{\mu\nu} =  \delta {g}_{\mu\nu} +{2A(y)\over H} 
(\nabla_\mu \nabla_\nu +H^2 \gamma_{\mu\nu})\varphi(x)\,.
\label{gaugetransform}
\eeq
Moreover, $(\square +4H^2)\varphi\sim T$, and a nonzero-trace 
source could excite a ghost and become  unstable. 
However, such a source also leads to the breakdown of 
the linearized approach adopted above. We turn to the 
discussions of these issues in Section 3.

%%%%%%%%%%%%%%%%%%%%%%%%%%%%%%%%%%%%%%%%%%%%%%%%%%

\section{Perturbations in the presence of sources}

%%%%%%%%%%%%%%%%%%%%%%%%%%%%%%%%%%%%%%%%%%%%%%%%%%

As we have already discussed in Sections 1 and 2, non-conformal sources 
could excite ghost in the linearized reduction of the theory
(see, however, discussions in Section 4). If this 
were all, the sources would start to radiate ghosts 
until they'd disappear (if the sources  carry no conserved charges), 
or would accrete some negative ``gradient energy'' to produce
an effective source that by itself is conformal. However, in all 
these considerations it is important to remember that non-conformal  
sources lead to the breakdown of the linearized 
approximation; they surround themselves with a nonlinear halo of 
a variable curvature that locally dominates over the self-accelerated 
background. The naive linearized solutions, that could contain instabilities, 
are not guaranteed to penetrate the halo. We will present 
an evidence that the linearized solutions are indeed spurious 
in this case. Dynamically, the halo screens the bare mass of the source,
and because of this, the process of its formation is energetically favorable.

\subsection*{Complete set of nonlinear equations}

{}To discuss these issues it is useful to start with the DGP 
action in the ADM formalism \cite {ADM} (see, e.g., 
\cite {Dick}, \cite {CedricMourad}):
\beq 
{S}=\, {\mpl^2\over 2} \int d^4x\,dy \sqrt{ g} \left (  R~\delta(y)+
{m_c\over 2} 
\,N\,\left (R \,+\,K^2 -K_{\mu\nu}K^{\mu \nu} \right )\right )\,.
\label{ADMaction}
\eeq
Equations of motion of the theory are obtained  by varying 
the action (\ref {ADMaction}) w.r.t. $g_{\mu \nu}$, $N$ and  $N_\mu $. 
Here we reiterate a subset of two equations, the  
junction  condition across the brane, and the $\{55\}$ 
equation that can be obtained by varying  the action 
w.r.t. $N$. The former reads as follows:
\beq\label{jc}
G_{\mu\nu} - m_c(K_{\mu\nu}-g_{\mu\nu}K)&=&T_{\mu\nu}/\mpl^2~,
\eeq
where $G_{\mu\nu}$ is the 4D Einstein tensor 
of the induced metric $g_{\mu\nu}$ and $T_{\mu\nu}$ is the 
matter stress tensor. The $\{55\}$ equation  takes the form
\beq
R=  \,K^2 - K_{\mu \nu}^2 \,.
\label{yy}
\eeq
Note that (\ref {jc}) is valid only at $y=0^+$ while 
(\ref {yy}) should be fulfilled for arbitrary $y$.
The terms with the extrinsic curvature contain
derivatives w.r.t. the extra coordinate as well as
the $\{55\}$  and $\{ \mu 5 \}$  components of the metric.
By expressing from (\ref {jc})
$K$ and $K_{\mu\nu}$ in terms of $R_{\mu\nu}$ and $T_{\mu\nu}$ 
we obtain the following expression:
\beq
m_c^2R&=&{1\over 3}\left(R^2-3R^2_{\mu\nu}\right)+{1\over\mpl^2}
\left(2R_{\mu\nu}T^{\mu\nu}-{1\over 3}RT\right)-
{1\over \mpl^4}\left(T_{\mu\nu}^2-{1\over 3}T^2\right)\,.
\label{master0}
\eeq
The remaining equations correspond to the $\{\mu 5\}$ and 
bulk $\{\mu\nu\}$ equations. The  $\{\mu 5\}$ equation for arbitrary $y$ 
(which we copy  here again for convenience) reads as follows:
\beq
\nabla_\mu K= \nabla^\nu K_{\mu\nu}\,.
\label{mu5}
\eeq
The covariant derivative in the above equation 
is the one for $g_{\mu\nu}$. At $y=0$ Eq. (\ref {mu5})  is 
trivially satisfied due to  (\ref {jc}).  For $y\neq 0$, 
(\ref {mu5}) sets the relation between $N_\mu$, $N$ and $g_{\mu\nu}$.  
Hence, (\ref {mu5}) gives a relation between these 
quantities for both $y=0$ and $y\neq 0$.

One can use the bulk $\{55\}$ equation (\ref {yy}) to determine 
$N$ in terms of $N_\mu$ and $g_{\mu\nu}$, and then use 
(\ref {mu5}) in order to express $N_{\mu}$ in terms of 
$g_{\mu\nu}$.  If so, there must exist one more equation which 
should allow to determine the bulk $g_{\mu\nu}$ itself. This is 
the bulk $\{\mu\nu\}$ equation, to which we turn now.
The latter can be written in a few different 
ways. For the case at hand, the formalism by Shiromizu, Maeda and Sasaki 
\cite {Sasaki} is most convenient.
In this approach, the bulk $\{\mu\nu\}$ equation and the junction 
condition can be combined to yield a  $\{\mu\nu\}$ equation 
at $y\to 0^+$. This gives a ``projection''
of the bulk $\{\mu\nu\}$ equation onto the brane. Since the bulk 
itself is empty in our case, 
this is the most restrictive form one can work with. 
The equation reads as follows \cite {Sasaki,Cedric5,Cedric6,Aliev}:
\beq
G_{\mu\nu} = {1\over m_c^2} B_{\mu\nu} -  E_{\mu\nu}\,,
\label{sasaki1}
\eeq
where 
\beq
B_{\mu\nu}&\equiv& - {\tilde G}_{\mu\alpha}{\tilde G}^{\alpha}_\nu +
{1\over 3}{\tilde G}^\alpha_\alpha   {\tilde G}_{\mu\nu}
+{1\over 2} g_{\mu\nu}  {\tilde G}_{\alpha \beta} {\tilde G}^{\alpha\beta} 
-{1\over 6} g_{\mu\nu} ({\tilde G}^{\alpha}_\alpha)^2\,,\nonumber  \\ 
{\tilde G}_{\mu\nu} &\equiv& { G}_{\mu\nu}-T_{\mu\nu}/\mpl^2\,,
\label{B}
\eeq
and where all quantities are taken at $y=0^+$.
$G_{\mu\nu}$ denotes the 4D Einstein tensor of the metric $g_{\mu\nu}$, 
$E_{\mu\nu}$ denotes the electric part of the bulk Weyl tensor, 
projected onto the brane. An important property of the Weyl 
tensor is that it is invariant under conformal 
transformations. Moreover, $E_{\mu\nu}$ is 
traceless.

%%%%%%%%%%%%%%%%%%%%%%%%%%%%%%%%%%%%%%%%%%%%%%%%%%
\subsection*{Perturbations in nonperturbative domain}

%%%%%%%%%%%%%%%%%%%%%%%%%%%%%%%%%%%%%%%%%%%%%%%%%%

One way in which the presence of instability was argued 
in the literature \cite {Ratt} is by looking at the scalar 
bending mode and observing that its kinetic term flips the sign 
due to curvature of the self-accelerated background.  
However, the scalar mode carries no gauge-independent 
physical information\footnote {This is true even in the so called 
decoupling limit where as was shown in \cite {GIde}, the scalar mode 
does not decouple in DGP from the tensor modes.}.
Because of this, no conclusion can be obtained just 
by looking at this mode; as we have seen in Section 2, 
the mixing with the tensor should be taken into account.

To emphasize once again how misleading 
the scalar mode could be, we briefly recall below 
a well-known example.   
Consider conventional 4D GR coupled to matter on a flat space. 
No instabilities are present in this theory. Let us perform now 
conformal transformation of the metric by the 
conformal factor $(1+\pi)$. The resulting Lagrangian 
contains the tensor field, kinetic mixing term 
of the tensor to $\pi$, and, most importantly,  
a wrong-sign (ghost-like) kinetic term for the scalar $\pi$.
Moreover, this scalar mode does couple to matter in the new frame.
If one simply ignores  the tensor mode (by putting it equal 
to zero, or by considering fields on which the conformally 
rescaled  $R$ is zero), one would naively conclude 
that $\pi$ is a ghost that  could be emitted by the source. 
However, this would be an erroneous conclusion, since we started 
with the theory with no ghost in the first place. The error is made by 
neglecting the mixing of $\pi$ with the tensor, and ignoring the 
fact that $\pi$ by itself carries no physical information.

The above example is instructive, and does certainly demonstrate the 
danger of drawing conclusions in gravity (as well as in gauge theories) 
based on the helicity-0 sector alone. However, the situation is not 
as simple in the DGP model.  This is because there are nonlinear 
mixing terms between the scalar and tensor modes 
which, on classical backgrounds,  give rise to 
quadratic  mixing terms between them. Because of this, as we will 
argue below, what determines the mixing 
at scales $r\lsim r_* $, is the local background  due to the source,
and not the global cosmological background\footnote{It is worth 
mentioning that the dynamics of small perturbations of the bending 
mode $\pi$ of Refs. \cite {Luty,Ratt},  in the nonperturbative regime, 
is completely dominated by the local background.  That is to say, 
the fluctuations of the $\pi$ field are 
well above the strong coupling scale of the theory,
the later being determined via  the graviton scattering amplitude 
on Minkowski background \cite {Luty}. However, the ``effective'' 
strong coupling scale  of the theory, even in the case of the graviton 
scattering on an empty space, is determined by the local non-perturbative 
background due to gravitons themselves, 
or due to any other supercritical sources which are present in a given
setup. This effective scale is much higher than the  strong coupling 
scale of \cite {Luty} (see, discussions in the first reference 
of \cite {GI}).}.

Let us consider small perturbations around a classical solution 
(e.g., such as the nonperturbative solutions of \cite{GI}) where there is 
significant curvature for distances smaller than a certain distance $r_*$.
Taking perturbations of the following form:
\beq
R_{\mu\nu}=R_{\mu\nu}^{\rm cl}+\delta R_{\mu\nu}~,
\eeq
we obtain from the first order expansion of (\ref{master0})
\beq\label{mast}
m_c^2\delta{R}&=&{2\over 3}R^{\rm cl}\delta R-
2 R^{\rm cl}_{\mu\nu} \delta  R^{\mu\nu} +
{2\over \mpl^2} R^{\rm cl}_{\mu\nu}T^{\mu\nu}-
{1\over 3\mpl^2} R^{\rm cl} T\,.
\eeq
Notice the crucial role played by the non-vanishing curvature of the classical 
solutions in determining the properties of the perturbations.

{}Let us first discuss the conventional branch. In this case, for distances 
much greater than $r_*$ the Ricci tensor is negligible. This yields 
$\delta  R=0$ 
(we consider 4D sources localized to a region smaller than $r_*$). While in the
regime of distances smaller than  $r_*$ we get
\beq\label{conv}
m_c^2\delta  R&=&{2\over 3} R^{\rm sch}\delta R-
2 R^{\rm sch}_{\mu\nu} \delta  R^{\mu\nu} +
{2\over \mpl^2} R^{\rm sch}_{\mu\nu} T^{\mu\nu}-
{1\over 3\mpl^2} R^{\rm sch} T\,,
\eeq
where the superscript $''{\rm sch}''$ refers to the nonzero 
curvature solution in the conventional branch. 

{}On the self-accelerated branch, in the large distance 
regime where the curvature is de Sitter like (with Hubble parameter equal to 
$m_c$) Eq. (\ref{mast}) reduces to $\delta  R=-2\delta  R$,
which amounts to $\delta R=0$ as in the conventional branch.
In the short distance regime the curvature  flips the 
sign with respect to the conventional branch\footnote{This statement is a 
direct consequence of the structure of the Israel junction conditions 
in DGP, and is independent of particular details of the solutions at 
hand.}:
\beq
R^{\rm sch}_{\mu\nu}\to -R^{\rm sch}_{\mu\nu}\,.
\eeq
Therefore, (\ref{mast}) yields, in this case
\beq
m_c^2\delta  R&=&-{2\over 3} R^{\rm sch}\delta R+
2 R^{\rm sch}_{\mu\nu} \delta  R^{\mu\nu} -
{2\over \mpl^2} R^{\rm sch}_{\mu\nu} T^{\mu\nu}+
{1\over 3\mpl^2} R^{\rm sch} T\,,
\label{conv1}
\eeq
with only the sign differences of the r.h.s. as compared to (\ref{conv}).

At scales below $r_*$ the local curvature $ R^{\rm sch}$ overcomes 
$m_c^2$, and becomes larger  and larger as one approaches the source
(still remaining much smaller that the magnitude of the Riemann
tensor components of the conventional 4D Schwarzschild metric 
\cite {GI}). Because of this, the expressions on the l.h.s. of 
(\ref{conv}) and (\ref{conv1}) can be neglected as compared to the 
terms on the r.h.s. In this approximation, 
the above two equations coincide up to an overall sign. 

Very similar calculations can be performed with a more general 
equation (\ref {sasaki1}). Since the conclusions of those 
studies are analogous to what we found above, and, moreover, the 
above presented expressions can 
be obtained by taking trace of (\ref {sasaki1}), 
we won't present those results here.

Therefore, the dynamics of small perturbations 
at scales $r\lsim r_*$ is completely determined by the local
metric due to the source, and not by the cosmological background.
The linearized perturbations, which by themselves are 
valid at $r\gsim r_*$, are not guaranteed to match upon the 
solutions inside the $r\lsim r_*$ domain. Physically, this is because the 
halo of a variable curvature that extends to scales $r\sim r_*$, 
screens the stress-tensor of the source \cite {GI}, 
making it hard for the naive linearized solutions to match it 
smoothly. For instance, in the static solution of \cite {GI}, 
the modes that at $r\gsim r_*$  match smoothly the interior solution,
are not the modes  of the naive linearized approximation. The latter 
are spurious.

\subsection*{Global constraints on self-accelerated background}

In this subsection we consider Euclidean continuation of the 
self-accelerated background and discuss the global constraints 
imposed on the linearized perturbations in a theory with, 
and without, sources.

We start with the constraints similar to the ones emerging
in 4D GR on a dS space-time  with a closed spatial section. 
This part of our  presentation is a straightforward  
generalization to one more dimension of the approach 
of Ref. \cite {Higuchi1}. Later in this subsection we will discuss 
the constraints that are more specific to the DGP model.
However, we should point out that, in terms of the Lorentzian
signature space-time, the results of the present section
are specific to a foliation of the dS space-time that covers the 
whole dS hyperboloid. We do not expect  similar results to arise for 
flat and open dS backgrounds in the absence of 
non-conformal sources. On the other hand, if the non-conformal 
sources were present, such global constraints could 
formally arise for different reasons. However, since in the latter 
case the linearization breaks down, these constraints should 
have no importance for the self-accelerated solution.

Consider the euclideanized dS space, $S^4$.
Suppose there exists some symmetric tensor $\pi^{\mu\nu}$
which is covariantly conserved
\beq
\nabla_\nu \pi^{\mu\nu} =0\,,
\label{pimunu}
\eeq
with $\nabla_\nu$ being the covariant derivative compatible with 
the metric $ g_{\mu\nu}$. Then, the integral below is a total derivative, 
and, assuming that fields are single-valued, it vanishes on the $S^4$
\beq
Q_\zeta \equiv \int_{S^{4}} d^4x \sqrt{g} \left 
( {\pounds}_\zeta g_{\mu\nu}\right )\pi^{\mu\nu} =0\,.
\label{Q}
\eeq
The Lie derivative  ${\pounds}_\zeta$ w.r.t. the vector field
$\zeta^\alpha$  is defined as follows
\beq
{\pounds }_\zeta t_{\mu\nu} &= & \zeta^{\alpha} 
\partial_\alpha t_{\mu\nu}+ t_{\mu\alpha}
\partial_\nu \zeta^{\alpha} + t_{\nu\alpha}
\partial_\mu \zeta^{\alpha} \nonumber \\
&= &\zeta^{\alpha} {\tilde \nabla}_\alpha t_{\mu\nu} + t_{\mu\alpha}
{\tilde \nabla}_\nu \zeta^{\alpha} + t_{\nu\alpha}
{\tilde \nabla}_\mu \zeta^{\alpha}\,,
\label{lie}
\eeq
where $ t_{\mu\nu}$ is an arbitrary smooth tensor 
and the second line follows from the first one given that 
${\tilde \nabla}_\nu$ is a covariant  derivative of an {\it arbitrary} 
metric whose affine connection is torsion-free
($g_{\mu\nu}$ is one example, but not the only one).

Let us split $g_{\mu\nu}$ and 
$\pi^{\mu\nu}$ in their background and perturbative parts
\beq
g_{\mu\nu}\equiv \gamma_{\mu\nu}+h_{\mu\nu} \,,~~~~ \pi^{\mu\nu}=
{\hat \pi}^{\mu\nu} + p^{\mu\nu}\,.
\label{splitgpi}
\eeq
Suppose that $\zeta^{\alpha}$ is a Killing vector of the background 
space-time, that is
\beq
{\pounds }_\zeta \gamma_{\mu\nu} =0\,.
\label{liegamma}
\eeq
Moreover, below  we will 
consider $\pi^{\mu\nu}$'s for which 
\beq
{\pounds }_\zeta {\hat \pi}^{\mu\nu} =0\,.
\label{liegamma1}
\eeq
Using these properties we  
expand the expression for $Q_\zeta$  (\ref {Q}) 
up to the second order in small perturbations. 
Ignoring the surface terms on the $S^4$, we get
\beq
Q_\zeta \simeq \int d^4x \sqrt{\gamma} \left 
( {\pounds }_\zeta h^{(1)}_{\mu\nu}\right )p^{(1)\mu\nu}=0\,.
\label{qsecond}
\eeq
Here $h^{(1)}$ and $p^{(1)}$ are the solutions of the first
order equations. Expression (\ref {qsecond}) is an additional 
constraint imposed on these perturbations. 

Let us now turn to the expression for  $\pi^{\mu\nu}$
that are specific to the model at hand. There are at least 
three tensors that satisfy the properties of $\pi^{\mu\nu}$ described above. 
These are:
\beq
K^{\mu\nu}- g^{\mu\nu}K,~~~~G^{\mu\nu}\,,~~~~~T^{\mu\nu}\,.
\label{tensors}
\eeq
Only two of these three are independent as they are 
related to each other by the junction condition
$G^{\mu\nu}- m_c(K^{\mu\nu}- g^{\mu\nu}K)= T^{\mu\nu}/\mpl^2$.
The global constraint (\ref {qsecond}) for these tensors  
reads respectfully:
\beq
Q^K_\zeta& =& \int d^4x \sqrt{\gamma} \left 
( {\pounds }_\zeta h^{(1)}_{\mu\nu}\right )
(K^{(1)\mu\nu}-  (g^{\mu\nu}K)^{(1)})=0\,, \label{Kglobal} \\
Q^G_\zeta &=& \int d^4x \sqrt{\gamma} \left 
( {\pounds }_\zeta h^{(1)}_{\mu\nu}\right )
G^{(1)\mu\nu}=0\,, \label {Gglobal} \\
Q^T_\zeta &=& \int d^4x \sqrt{\gamma} \left 
( {\pounds }_\zeta h^{(1)}_{\mu\nu}\right )
T^{(1)\mu\nu}=0\,. \label {Tglobal} 
\eeq
Any one of the above constraints  follows 
from the other two. Let us focus on the last one (\ref {Tglobal}).
In an expanded form it reads:
\beq
\int d^4x \sqrt{\gamma} \left ( \zeta^{\alpha} 
\nabla^{(\gamma)}_\alpha h^{(1)}_{\mu\nu}  + h^{(1)}_{\mu\alpha}
\nabla^{(\gamma)}_\nu \zeta^{\alpha} + h^{(1)}_{\nu\alpha}
\nabla^{(\gamma)}_\mu \zeta^{\alpha}  \right ) T^{(1)\mu\nu}=0\,,
\label{Tgloballong}
\eeq
where $\nabla^{(\gamma)}_\alpha$ denotes the covariant derivative 
compatible with the background metric $\gamma_{\mu\nu}$.
The significance of this constraint is the following.
one would calculate 
$h^{(1)}$ in terms of the $\nabla^{(\gamma)}_\alpha$
and an arbitrary source $T^{(1)}$ (as it is done in the next section).
The question whether the obtained $h^{(1)}$ can or cannot 
be continued into the full nonlinear theory would have 
been decided based on whether or not it could satisfy the constraint  
(\ref {Tgloballong}). The latter, in general, is hard to be fulfilled
automatically, and it would impose a constraint on the 
acceptable sources $T^{(1)}$. However, in the case of the self-accelerated 
solution with the sources the situation is slightly different. 
Local curvature within an arbitrary overdensity above the background is 
bigger than $m_c^2$.  As a result, the perturbative expansion 
should break down for all these sources. Because of this, the 
derivation of (\ref {Tgloballong}) presented above fails. 
The effects of the nonperturbative domain lead
to screening  of the mass of the source \cite {GIde}. 

In the remainder of this subsection, we would like to discuss the 
global constraints on perturbations of the self-accelerated 
background without sources. These are more specific to the DGP model,
as they are based on a nonlinear trace equation that is very different 
from that of GR. However, as we mentioned at the beginning of this 
subsection, these constraints are specific to the foliation of the 
dS space-time that covers the whole dS hyperboloid. 

To get to the point as quickly as possible
we again turn to the euclideanized version of the background 
dS space with the topology of a four-sphere $S^4$, 
and  look at one of the exact  equations of motion 
of the DGP model 
\beq
3 m_c^2R\,=\, R^2-3R^2_{\mu\nu}\,.
\label{master0011}
\eeq
This equation is valid at the position of the brane 
in an arbitrary coordinate system. We will be expanding this equation 
about the background $S^4$ up to the second order in perturbations 
(which corresponds to the expansion up to the cubic  
order in the action). Hence, 
\beq
g_{\mu\nu} = \gamma_{\mu\nu} + \delta g_{\mu\nu}\,,~~~g_{\mu\nu}=
\gamma^{\mu\nu}- \delta g^{\mu\nu} + 
\delta g^{\mu\alpha} \delta g_{\alpha}^{\nu}+...,
\label{metricexp}
\eeq
where the first equation in (\ref {metricexp})  
is just the definition of $\delta g_{\mu\nu}$.
Furthermore, we use the following standard notations for the 
expansion of the Ricci tensor: 
\beq
R_{\mu\nu} = 3H^2 \gamma_{\mu\nu} + R^{(1)}_{\mu\nu}+ R^{(2)}_{\mu\nu}\,,
\label{ricciexp}
\eeq
where $R^{(1)}_{\mu\nu}$  and $R^{(2)}_{\mu\nu} $
denote  respectively the linear and quadratic in perturbations 
terms (for convenience, the explicit expressions for these quantities 
are given in Appendix C, see also \cite {Wald}). 
It is also convenient to adopt the following notations:
\beq
\delta g_{\mu\nu} =h^{(1)}_{\mu\nu}+ h^{(2)}_{\mu\nu}+...,
\label{metrexp12}
\eeq
where $ h^{(1)}_{\mu\nu} $  denotes  a solution of the first order 
equation, while $ h^{(2)}_{\mu\nu} $ is a solution of the  second 
order equations. The decomposition of  (\ref {metrexp12}) is 
useful for keeping  track of terms in the corresponding orders. For instance,
the term $R^{(1)}_{\mu\nu}(h^{(1)})$ is a first order quantity, while
the terms  $R^{(1)}_{\mu\nu}(h^{(2)})$, $R^{(2)}_{\mu\nu}(h^{(1)})$ and 
$h^{(1){\mu\nu}} R^{(1)}_{\mu\nu}(h^{(1)})$ are all 
the second order quantities. On the other hand, the terms 
$h^{(2){\mu\nu}} R^{(1)}_{\mu\nu}(h^{(1)})$, 
$h^{(1){\mu\nu}} R^{(2)}_{\mu\nu}(h^{(1)})$ appear in the third order,
while $R^{(2)}_{\mu\nu}(h^{(2)})$ in the fourth order. 
We will be ignoring all the terms beyond the second order.

To proceed further we need to specify the gauge conditions at each order.
In the first order we choose the  gauge 
\beq
\gamma^{\mu\nu} h^{(1)}_{\mu\nu} =0\,.
\label{ttpertgauge}
\eeq
The first order source free equations of motion also allow to 
set  (although the latter is not necessary for our derivations)
\beq
\nabla^\mu h^{(1)}_{\mu\nu} =0\,.
\label{ttpertgauge1}
\eeq
In the second order, however, we impose only the tracelessness condition
\footnote{This does not fix the gauge freedom in the second order completely, 
but is enough for our purposes, i.e., our considerations apply irrespective 
of an additional gauge fixing condition that one may need 
to introduce to fix the residual gauge freedom 
of the second order perturbations, as long as that conditions is consistent
itself. Notice that we do not impose the covariant 
transversality condition  in the second order, as this would have 
over-constrained the system.}:
\beq
\gamma^{\mu\nu} h^{(2)}_{\mu\nu} =0\,.
\label{tpertsecond}
\eeq
Now we are ready to proceed to the expansion of 
(\ref {master0011}).  Since $m_c=H$ and 
$\gamma^{\mu\nu}R^{(1)}_{\mu\nu}(h^{(1)})=0$ , 
we find in the second order in perturbations:
\beq
R\equiv g^{\mu\nu} R_{\mu\nu}  
=12H^2 + \gamma^{\mu\nu}R^{(1)}_{\mu\nu}(h^{(2)}) + 
\gamma^{\mu\nu}R^{(2)}_{\mu\nu}(h^{(1)}) \nonumber \\
- h^{(1){\mu\nu}} R^{(1)}_{\mu\nu}(h^{(1)})+ 3H^2(h^{(1)}_{\mu\nu})^2\,.
\label{Rpert}
\eeq
The terms on the r.h.s. of (\ref {master0011}) 
are expanded to the second order as follows:
\beq
R^2 = 144H^4 + 24 H^2 \gamma^{\mu\nu}R^{(1)}_{\mu\nu}(h^{(2)})
+ 24 H^2 \gamma^{\mu\nu}R^{(2)}_{\mu\nu}(h^{(1)}) \nonumber \\
-  24 H^2 h^{(1){\mu\nu}} R^{(1)}_{\mu\nu}(h^{(1)}) + 72 H^4
(h^{(1)}_{\mu\nu})^2\,,
\label{R2pert}
\eeq
and 
\beq
R^2_{\mu\nu}\equiv  R_{\mu\nu}R_{\alpha\beta}g^{\mu\alpha}g^{\nu\beta}=
36H^4 + 6 H^2 \gamma^{\mu\nu}R^{(1)}_{\mu\nu}(h^{(2)})
+ 6 H^2 \gamma^{\mu\nu}R^{(2)}_{\mu\nu}(h^{(1)}) \nonumber \\
- 12 H^2 h^{(1){\mu\nu}} R^{(1)}_{\mu\nu}(h^{(1)}) + 27 H^4
(h^{(1)}_{\mu\nu})^2 + ( R^{(1)}_{\mu\nu}(h^{(1)}) )^2\,. 
\label{Rmunu2pert}
\eeq
Substituting the expressions (\ref {Rpert}), (\ref {R2pert})
and (\ref {Rmunu2pert}) into the equation (\ref {master0011}) we get
\beq
3H^2 \gamma^{\mu\nu}R^{(1)}_{\mu\nu}(h^{(2)})+ 
3H^2 \gamma^{\mu\nu}R^{(2)}_{\mu\nu}(h^{(1)})+ 15H^2 
h^{(1){\mu\nu}} R^{(1)}_{\mu\nu}(h^{(1)})\nonumber \\
 -18 H^4 (h^{(1)}_{\mu\nu})^2
-3 (R^{(1)}_{\mu\nu}(h^{(1)}) )^2=0\,.
\label{integrand11}
\eeq
All the terms in the above equation are of the second order. 
This equation 
should be used to determine the second order metric perturbation 
$h^{(2)}_{\mu\nu}$ in terms of the squares of the first order 
perturbation  $h^{(1)}_{\mu\nu}$ and its derivatives. However, there 
is more to it: Only the first term on the l.h.s. of  (\ref {integrand11}) 
contains the second order metric perturbation 
$h^{(2)}_{\mu\nu}$, the rest contains only the first order metric
$h^{(1)}_{\mu\nu}$. Moreover, since  $h^{(2)}_{\mu\nu}$ is traceless,
then $\gamma^{\mu\nu}R^{(1)}_{\mu\nu}(h^{(2)})= \nabla^\mu 
\nabla^\nu h^{(2)}_{\mu\nu}$.
As a result, the first term on the l.h.s. of (\ref {integrand11}), which 
also happens to be the only term containing $h^{(2)}_{\mu\nu}$, 
is a total derivative. Therefore, if we integrate both sides of 
(\ref {integrand11}) w.r.t. the background space, which in our case
in $S^4$, the term containing $h^{(2)}_{\mu\nu}$ will 
drop out as a total derivative. The resulting integral equations reads:
\beq
\int_{S^4}d^4x\sqrt{\gamma} 
\left (H^2 \gamma^{\mu\nu}R^{(2)}_{\mu\nu}(h^{(1)}) + 5H^2 
h^{(1){\mu\nu}} R^{(1)}_{\mu\nu}(h^{(1)})- 6 H^4 (h^{(1)}_{\mu\nu})^2 -
(R^{(1)}_{\mu\nu}(h^{(1)}) )^2 \right )\nonumber \\ =0\,. 
\label{global1}
\eeq
This equation contains only  the first order perturbations
$h^{(1)}_{\mu\nu}$, and represents an {\it additional} constraint 
on these perturbations imposed by the second order equations.
This constraint cannot be captured in the linearized theory.
Although we derived (\ref {integrand11}) using a particular gauge,
it represents an integral of a gauge-invariant equation 
(\ref {master0011})\footnote{In conventional GR on $S^4$ 
Eq. (\ref {Rpert}) can be used to derive a global constraint.}. 

Can the first order perturbations that are solutions of the linearized 
equations  satisfy the above constraint? To answer this question  we 
use the explicit expressions for the quantities in (\ref {global1}) 
(see, e.g., Appendix C), which after integration by parts yields:
\beq
\int_{S^4}d^4x\sqrt{\gamma}  \left((\square h^{(1)}_{\mu\nu})^2 + 5H^2 
(\nabla_\alpha h^{(1)}_{\mu\nu})^2 \right)=0\,.
\label{global2}
\eeq
The integrand on the l.h.s. of (\ref {global2})
is a positive semi-definite quantity in the Euclidean space 
that we are dealing 
with. Hence, the global constraint (\ref {global2}) implies 
that $\square h^{(1)}_{\mu\nu}=0$, and $ \nabla_\alpha h^{(1)}_{\mu\nu}=0$. 
These can only be satisfied by trivial solutions! 

Summarizing, the nonlinear equations impose very severe 
constraints on euclideanized linear perturbations. If fact, on a brane with no 
matter sources considered above, no non-trivial  perturbations appear 
to survive these constraints. Then what is physics of perturbations
on such a background? 
The answer to this question, as we discussed before,  
lies in non-perturbative dynamics. It so happens that realistic 
sources, whether they are due to nonlinear self-interactions of gravity 
itself, or external matter, give rise to the breakdown of the naive linearized 
approximation used in this section. This breakdown  obfuscates the meaning 
of the conventional linearization  approximation and constraints discussed 
above. On the other hand, it also  suggests  that the required calculations 
should be done using different approximations, e.g., the one 
we briefly discussed in Section 1. 

 \section{Truncated theory with sources}

In this Section we consider the self-accelerated background  
on which the theory is  truncated to the linearized level. 
As we have discussed before,
such a theory is ruled out already by the Solar system data
because of the vDVZ discontinuity. Yet, this is the model based on which 
the conclusions of the instability of the self-accelerated background 
have been reached in the literature. 
As we argued in the previous Sections, the only model that makes
physical sense to discuss is the one in which non-linearities are retained, 
whereas the truncated model has only a marginal interest. Nevertheless, it 
can be used to discuss certain theoretical questions,
such as its relation to the  model of massive gravity with the 
enhanced symmetry \cite {Deser}. However, even in this case, 
there are subtle issues concerning the analytic structure and boundary 
conditions of the Green's functions, which have not be clarified.
This issues are important for determining the 
linearized spectrum of the theory, as we 
will discuss in detail in the present section.

%%%%%%%%%%%%%%%%%%%%%%%%%%%%%%%%%%%%%%%%%%%%%%%%%%%%%%%%%%%%%%%%%%%%%%

\subsection*{The square root and boundary conditions}

%%%%%%%%%%%%%%%%%%%%%%%%%%%%%%%%%%%%%%%%%%%%%%%%%%%%%%%%%%%%%%%%%%%%%%

To get started, let us pick a coordinate 
system in which the brane is straight
and located at the $y=0$ point in the fifth dimension. 
Let us look at the trace of the 4D Einstein equation at the brane 
without  matter or radiation
\beq
R=3 m_c K\,.
\label{trace00}
\eeq
Both sides of (\ref {trace00}) are  calculated at $y=0^+$.
The key point is that $K$ can take negative or positive values
depending on the boundary conditions in the bulk. This could be seen, e.g.,
in a gauge in which $K_{\mu\nu}\propto \partial_y g_{\mu\nu}(x,y)$,
where $g_{\mu\nu}(x,y)$ denotes the $\{\mu\nu \}$ components of the 5D 
metric. If the metric falls off the brane in the $y$ direction, then the 
r.h.s. of (\ref {trace00}) could be negative. We call this the 
$(-)$ branch. Alternatively, if the metric grows off the brane 
in the $y$ direction, the r.h.s. of (\ref {trace00}) could be 
positive. We call this the $(+)$ branch. 
Clearly, both the $(-)$ and $(+)$ branches admit Minkowski 
space-time as a solution of (\ref {trace00}). However, the 
key point is that the $(+)$ branch, in addition, admits the 
self-accelerated solution.

To get more insight into this set of issues,
let us rewrite (\ref {trace00}) as a Friedmann equation  
for a spatially-flat  background \cite {Cedric}.
It is instructive \cite {G} to discuss the properties of the $(-)$ and $(+)$ 
branches in terms of this equation. For the $(-)$  branch it takes the form:
\beq
H^2+m_c|H|=0\,, 
\label{-}
\eeq
where $H$ denotes the Hubble parameter as seen 
by a 4D brane observer. The solution $H=0$ corresponds 
to a Minkowski background of the  $(-)$  branch. Small 
perturbations about this space are stable \cite {DGP}. 

For the  $(+)$ branch, on the other hand, 
the Friedman equation reads:
\beq
H^2\,-\,m_c\,|H|\,=\,0\,.
\label{F2}
\eeq
There is a set of two solutions. The solution $(A_+)$ 
with $H=0$, and $(B_+)$ with $|H|=m_c$.
The $(A_+)$ solution is just a Minkowski background 
of the  $(+)$ branch.
The solution $(B_+)$, that is called the 
self-accelerated solution \cite {Cedric}, is the most interesting one,
as it describes an accelerated expansion of the Universe. 

In addition to the freedom in choosing 
boundary conditions in the bulk there is also a freedom in 
setting a direction of the time arrow $t$ on the brane.
This freedom is reflected in the above equations 
as a symmetry $H\to -H$.
This suggests that a solution with a positive (negative) 
$H$ could be expanding (contracting) for one choice of 
the direction of $t$ and contracting (expanding) 
for the opposite choice.

Some of the above described properties of the 
$(-)$ and $(+)$ branches can also be seen  by analyzing the 
perturbative amplitudes on a {\it flat} space. For them,  
the operator $\partial_y$ in the expression for $K_{\mu\nu}$
is replaced by the square root of 
the d'Alembertian $\pm \sqrt {-\square_4}$. A choice
of the bulk boundary conditions  is in one-to-one correspondence with  
the choice of  the sign of $\pm \sqrt {-\square_4}$. Furthermore, 
the choice of this  sign accompanied with the choice 
of the direction of the time arrow 
has to do with the causality  and unitarity of the 
amplitudes; it determines whether there could or 
could not be tachyon and ghost instabilities in the linearized 
theory. Since this is an important  issue, and has direct relevance to our 
calculations, we will reiterate and sharpen 
below some of the discussions of Ref. \cite {G}.

For the $(-)$ branch, the one-particle  exchange amplitude 
for two very weak sources on the brane 
(in the momentum space w.r.t. the four 
worldvolume coordinates) takes the form \cite {DGP}:
\beq
{\cal M}^{(-)}(p,\,y)\,=\,{T^2_{1/3}\,{\rm exp }\left 
(- \sqrt{p^2} |y| \right ) \over p^2\,+\,m_c\,\sqrt{p^2}}\,,
\label{A13}
\eeq
where we denote the Euclidean four-momentum squared by 
$ p^2\equiv -p_0^2+p_1^2+ 
p_2^2+p_3^2\equiv
p_4^2 + p_1^2 + p_2^2 +p_3^2$,
and
\beq
T^2_{1/3} \,\equiv\, 8\,\pi\,G_N \left ( 
T^2_{\mu\nu}\,-\,{1\over 3}\,T^\alpha_\alpha \cdot T^\beta_\beta \right 
)\,.
\label{T1third}
\eeq
The amplitude (\ref {A13}) was constructed 
by imposing the decreasing  boundary conditions in the 
$y$ direction for positive  {\it Euclidean} momenta. In 
the Lorentzian signature this corresponds to an oscillatory 
dependence on the $y$ coordinate with a plus sign of the exponent 
${\cal M}^{(-)}\sim {\rm exp}(+i \sqrt{-p^2}|y|)$. In the physical domain, 
$-p^2\ge 0$, this corresponds to a retarded amplitude for a given 
choice of the time direction.
Furthermore, there is a branch-cut on the complex plane of 
$-p^2$ on the positive real semi-axes 
due to the square root in (\ref {A13}),
and the pole at $\sqrt{p^2}= -m_c$ is on a second nonphysical 
Riemann sheet (see, \cite {G} for detailed 
discussions.)  This pole has a positive residue, 
while the residue of the pole at  $p^2=0$ is zero.
Thus, it describes a metastable graviton without 
ghost or tachyon pathologies.

For the $(+)$ branch, on the other hand, the amplitude is determined by 
choosing  the other sign of the square root.  
As a result, one obtains a Euclidean amplitude that grows with $y$
\beq
{\cal M}^{(+)}(p,\,y)\,=\,{T^2_{1/3}\,{\rm exp}\left 
( \sqrt{p^2} |y| \right ) \over p^2\,-\,m_c\,\sqrt{p^2}}\,.
\label{A134}
\eeq
In  the Lorentzian signature this corresponds to 
an oscillatory  solution  with a negative exponent
${\cal M}^{(+)}\sim {\rm exp} (-i \sqrt{-p^2} |y|)$,  
and describes  an advanced amplitude. 
The expression (\ref {A134}) differs from  (\ref {A13}) 
also by the position of the pole in the denominator. The pole
at $\sqrt{p^2} =m_c$ is located on a first Riemann sheet
on a negative real semi-axes of $-p^2$, hence this pole is tachyonic.
This suggests that the space-time flat solution of the $(+)$ branch 
is unstable (this instability could be stronger than tachyonic)!
However, the situation can be reversed by replacing 
the time coordinate  $t$, 
by $-t$ on the $(+)$ branch. Only in this case the Minkowski solution
of the $(+)$ branch has a consistent 
physical interpretation.  After this replacement, (\ref{A134}) will 
describe a retarded amplitude for the new choice of the time direction. 
Moreover, the tachyonic pole becomes a resonance type pole. This is simply
because (\ref{A134}) turns identically into (\ref{A13}). 

On the self-accelerated background of the $(+)$ branch one can calculate 
two amplitudes with the behavior similar to (\ref{A13})  and (\ref{A134}), 
as we will show in this  Section. However, unlike 
the flat-space example discussed above, the replacement  $t\to -t$
does not transform one amplitude into the other one.

Summarizing, there are two choices to be made, one for the direction of $y$
and another for $t$. On a Minkowski background these two choices are 
degenerate. However, on a curved background the degeneracy is lifted and,
in general, one should study each particular case separately, 
as it will be done in this Section.

\subsection*{Spectrum in the presence of a source}

Let us introduce a source $T_{AB}(x,y)= \delta^\mu_A 
\delta^\nu_B T_{\mu\nu}(x)\delta (y)$.  In the gauge that we used
in Section 2, the source would bend the brane \cite {GT}.
Hence, we   define a new coordinate system
$({\tilde x}, {\tilde y})$ in which the brane is straight, and 
impose the following   gauge fixing conditions 
\beq
\nabla^\mu \delta {\tilde g}_{\mu\nu} = \nabla_\nu {\bar g}^{\alpha\beta}
\delta {\tilde g}_{\alpha\beta}\,,~~~ \delta {\tilde g}_{55}= 
\delta {\tilde g}_
{\mu5} =0\,.
\label{tttgauge}
\eeq
The new and old coordinate systems, and fields,  
are related by 5D gauge transformations (\ref {gaugetransform}). 
In terms of the new variables the metric is written as
\beq
ds^2 = \left ( {\bar g}_{\mu\nu}+ \delta {\tilde g}_{\mu\nu} \right )
d{\tilde x}^\mu d{\tilde x}^\nu + d{\tilde y}^2\,. 
\label{metricgt}
\eeq
Let us  look at the equations of motion.
The off-diagonal equation (\ref {mu5new})  written in the 
new coordinate system 
is identically satisfied. On the other hand, the $\{55\}$ equation (\ref {55}) 
takes the form:
\beq
-3n^2 {\bar g}^{\mu\nu}\delta {\tilde g}_{\mu\nu} = 6n \left (
{\bar g}^{\mu\nu}\delta {\tilde K}_{\mu\nu} - n{\bar g}^{\mu\nu}\delta 
{\tilde g}_{\mu\nu}\right )\,,
\label{new55}
\eeq
 which can be solved to obtain
\beq
{\bar g}^{\mu\nu}\delta {\tilde K}_{\mu\nu} = {n\over 2} 
{\bar g}^{\mu\nu}\delta {\tilde g}_{\mu\nu}\,.
\label{solv55}
\eeq
The latter is consistent with (\ref {gaugetransform}).

The junction  condition, taken at $0^+$,  reads as follows:
\beq
-{1\over 2} \left (\square_4 \delta {\tilde g}_{\mu\nu}  - 
\nabla_\mu\nabla_\nu \delta {\tilde g}^{\alpha}_{\alpha}
 \right ) + 2 H^2 \delta {\tilde g}_{\mu\nu}
-{H^2\over 2}\gamma_{\mu\nu} \delta {\tilde g}^{\alpha}_{\alpha} 
-H \left (  \delta {\tilde K}_{\mu\nu}- \gamma_{\mu\nu} 
\delta {\tilde K}^{\alpha}_{\alpha}\right ) 
=T_{\mu\nu}\,.
\label{munujunctiontilde}
\eeq
The first four terms on the l.h.s. of (\ref {munujunctiontilde})
is what appears in a 4D massive gravity at the point of enhanced 
symmetry (see Section 2 and Appendix B). Here, however,  we get 
additional terms due to the extrinsic curvature 
(the last two terms on the l.h.s.). This makes the self-accelerated 
background 
of the present model different from the massive 4D gravity at  
the special point \cite {Gorbunov}. Because of this, in the present model a 
non-conformal matter stress-tensor is allowed. Indeed, the trace 
of the junction equation  (\ref {munujunctiontilde})  takes the form:
\beq
3m_c {\bar g}^{\mu\nu} \delta {\tilde K}_{\mu\nu} =T\,.
\label{juncttracetilde}
\eeq
(The l.h.s. of this equations would have been zero for a 
massive 4D gravity at the special point (\ref {mR}), 
see Appendix B).

One  way to solve for the induced  metric 
$\delta {\tilde g}_{\mu\nu}$ is, 
to use (\ref {gaugetransform}), go to the variables 
without the tilde sign, and solve first for them. For 
$\delta {g}_{\mu\nu}$ we find the following equation
(valid at $0^+$) 
\beq
-{1\over 2} \left ( \square_4 -4 H^2      
 +H \partial_y \right ) \delta {g}_{\mu\nu} =
T_{\mu\nu} -{1\over 3} \gamma_{\mu\nu} T + 
\Pi_{\mu\nu} {T\over 3(\square_4 +4H^2)} \,,
\label{solmunu}
\eeq
where $\Pi_{\mu\nu} \equiv (\nabla_\mu \nabla_\nu +H^2 \gamma_{\mu\nu})$,
and we used $\delta {\tilde K}_{\mu\nu} = \delta {K}_{\mu\nu}+
\Pi_{\mu\nu} \varphi$. Here we note that the last term on the 
r.h.s. of (\ref {solmunu})  is somewhat alarming. 
It already has a pole (all operators are acting from left to right).
As a result,  the expression
for $\delta {g}_{\mu\nu}$ may end up having  a double pole. We will see
whether this potential problem can be avoided in the expression 
for the induced metric $\delta {\tilde g}_{\mu\nu}$.

The trace of the junction condition (\ref {juncttracetilde}) 
determines  $\varphi$:
\beq
\varphi(x) = {T\over 3 m_c (\square_4 +4H^2)}\,.
\label{solvPhi}
\eeq
We pause here. The pole in the above expression for $\varphi$ 
is tachyonic, it corresponds to a scalar of a negative mass 
square $m^2_t =-4H^2$.  
One could conclude that the brane bending mode $\varphi$ 
is at least as bad as a tachyon.
However, $\varphi$ is gauge dependent, and, in order to reach a 
conclusion one needs to calculate a gauge invariant 
physical amplitude.  

Finally, for future reference we write the 
expression that encompasses both the bulk and junction equations
for $\delta { g}_{\mu\nu} $:
\beq
\left (H \partial_y^2 + \left \{ {H\over A^2} + 2\delta (y) 
\right \} (\square_4
-4H^2) \right ) \delta { g}_{\mu\nu} =  
-4 \left ( T_{\mu\nu} -{1\over 3} \gamma_{\mu\nu} T 
+ H\Pi_{\mu\nu} \varphi \right ) \delta (y)\,.
\label{master00}
\eeq
We will use the above equation for the determination of the 
induced metric in the next subsection. The reminder of this 
subsection discusses a certain  subtlety that emerges in 
the above equation. A reader who is not interested in these 
details could skip to the next  subsection, where the 
actual calculations are performed.

This concerns the last term in the parenthesis 
on the r.h.s. of (\ref {master00}) (we will call it $\Delta_{\mu\nu})$:
\beq
\Delta_{\mu\nu}\equiv 
\left (\nabla_\mu\nabla_\nu + H^2\gamma_{\mu\nu}\right ) 
{1\over  \square_4 + 4H^2 } {T\over 3}\,.
\label{DeltaB}
\eeq
Looking at (\ref {DeltaB}) one may conclude  that
it has a tachyonic pole at $\square_4 + 4H^2 =0$. However, 
this would be a misleading conclusion. 
To see this we use the identity
\beq
(\square_4 - 4H^2)  ( \nabla_\mu \nabla_\nu + 
H^2 \gamma_{\mu\nu})({\rm scalar})  = (\nabla_\mu \nabla_\nu - 
H^2 \gamma_{\mu\nu}) (\square_4 + 4H^2)({\rm scalar})\,, 
\label{master}
\eeq
which allows us to rewrite $\Delta_{\mu\nu}$ as:
\beq
\Delta_{\mu\nu}\equiv {1\over  \square_4 - 4H^2 } 
\left (\nabla_\mu\nabla_\nu -H^2\gamma_{\mu\nu}\right ) 
{T\over 3}\,.
\label{Delta}
\eeq
As it is clear from (\ref {Delta}), the expression for  $\Delta_{\mu\nu}$
seems now  to have a pole at $\square_4 - 4H^2 =0$, for the operators 
that multiply it from the left. We will carefully deal with 
such subtleties in the next subsection.

\subsection*{Poles and residues}

In order to find the physical spectrum of the theory 
one should  look at the poles and residues of the 
induced metric:
\beq
\delta {\tilde g}_{\mu\nu}(x, y) = \delta g_{\mu\nu}(x, y) +
{2 A\over H} \Pi_{\mu\nu} \varphi \,.
\label{indfin}
\eeq
Substituting (\ref {indfin}) into  (\ref {solmunu}),
and using the identity (\ref {master})
we find the equation 
\beq
-{1\over 2} \left ( \square_4 -4 H^2      
 +H \partial_y \right ) \delta {\tilde g}_{\mu\nu} =
T_{\mu\nu} -{1\over 3} \gamma_{\mu\nu} T 
- \left (\nabla_\mu\nabla_\nu  -H^2\gamma_{\mu\nu}\right )   
{T\over 3H^2} \,. 
\label{tildemunu}
\eeq
As before, this equation is written for ${\tilde y}=0^+$. 
Note, that the troublesome double-pole structure that was present in  
(\ref {solmunu}), has remarkably canceled out in the above 
equation.

In order to find the expression for $\delta {\tilde g}_{\mu\nu}$
that would capture both the bulk and junction equations, 
we use (\ref {indfin}) in (\ref {master00}).  
The resulting equations for $\delta {\tilde g}_{\mu\nu} $
reads
\beq\label{master1}
\left (H \partial_y^2 + \left \{ {H\over A^2} + 2\delta (y) 
\right \} (\square_4
-4H^2) \right ) \delta {\tilde g}_{\mu\nu} = \nonumber \\  
-4 \left ( T_{\mu\nu} -{1\over 3} \gamma_{\mu\nu} T 
- \left (\nabla_\mu\nabla_\nu  -H^2\gamma_{\mu\nu}\right )   
{T\over 3H^2} 
\right ) \delta (y)\,
+{2\over A} \left (\nabla_\mu\nabla_\nu  -H^2\gamma_{\mu\nu}\right )
{T\over 3H}~.
\eeq
Here we would like to emphasize 
that  the last term on the r.h.s. of 
(\ref {master1})  does not vanish outside of the brane.
Therefore, even a source  that is  localized  
on the  brane, gives rise to  an  {\it effective} source
that is {\it delocalized} all over the bulk. It is the effective 
source that excites the physical metric 
$\delta {\tilde g}_{\mu\nu}$. Moreover, this effective source 
{\it grows} in the bulk (this is because in physical amplitudes 
the last term on  the r.h.s. of (\ref {master1}) should be multiplied by an
additional  factor of $A^2$ arising from the norm of the inner product).
This rather unusual behavior is a consequence of  the fact that 
in the coordinate system in which the brane is static and 
straight the background itself  grows  in the bulk as $A^2$. 

Let us now solve (\ref {master1}). It yields two independent solutions
\beq
\frac{1}{2} \delta \tilde{g}_{\mu \nu\pm} &=&  \frac{1}{{\cal O}^{(4)}} 
A^{\alpha_{\pm}\left({{\cal O}^{(4)}}\right)}
 \frac{1}{{\cal O}^{(4)} - H^2 \alpha_{\pm}\left({{\cal O}^{(4)}}\right)}
\left\{{\cal O}^{(4)}\left(T_{\mu \nu}- \frac{1}{3} \gamma_{\mu \nu} T\right) 
- \Pi^-_{\mu \nu} \frac{T}{3} \right\} \nonumber  \\&&
-\frac{1}{3 H^2} \frac{A}{{\cal O}^{(4)}} \Pi_{\mu \nu}^- T, \label{FORMALSOL}
\eeq
where ${\cal O}^{(4)}$, $\Pi_{\mu \nu}^-$ and $\alpha_{\pm}$ are respectively 
defined by 
\beq
{\cal O}^{(4)} &=& - \square_4+4H^2 ~,\\
\Pi_{\mu \nu}^- &=& \nabla_\mu \nabla_\nu - H^2 \gamma_{\mu \nu} ~,\\
\alpha_{\pm}\left(z\right) &=& \frac{1}{2}\left(1 \pm 
\sqrt{1+\frac{4 z}{H^2}}\right) ~.
\eeq
It is important to point out that the r.h.s. of (\ref{FORMALSOL}) has 
potentially two poles: the first one for the zero eigenvalues of 
${\cal O}^{(4)}$ and the 
second for zero eigenvalues of ${\cal O}^{(4)} - 
H^2 \alpha_{\pm}\left({{\cal O}^{(4)}}\right)$. 
These poles and their residues will be discussed in detail below.
%
%
%
%%%%%%%%%%%%%%%%%%%%%%%%%%%%%%%%%%%%%%%%%%%%%%%%%%%%%%%%%%%%%%%%%%%%%%%%

Equations (\ref{FORMALSOL}) can be rewritten as
\begin{eqnarray}\label{delg}
{1\over 2}\delta\tilde g_{\mu\nu\pm}&=& 
K_\pm({\cal O}^{(4)})\left[T^{TT}_{\mu\nu}+
{1\over 3}\left(P_{\mu\nu}{1\over {\cal Q}^{(4)}}-
{1\over {\cal O}^{(4)}}P_{\mu\nu}\right)
T\right]\nonumber\\&&
-{A\over 3H^2{\cal O}^{(4)}}P_{\mu\nu}T+{A\over 12 H^2}\gamma_{\mu\nu}T~,
\end{eqnarray}
where $P_{\mu\nu}=\nabla_\mu\nabla_\nu-{1\over 4}\gamma_{\mu\nu}\square_4$ 
({\it i.e.}, $\Pi^-_{\mu\nu}=P_{\mu\nu}-{1\over4}
\gamma_{\mu\nu}{\cal O}^{(4)}$), 
${\cal Q}^{(4)}=-\square_4-4H^2={\cal O}^{(4)}-8H^2$,
\begin{equation}
K_\pm(z)={A^{\alpha_\pm(z)}\over z-H^2\alpha_\pm(z)},
\end{equation}
and the decomposition of a covariantly conserved symmetric tensor in terms of 
its transverse-traceless, spin-0 longitudinal, and pure trace parts was
used, namely,
\begin{eqnarray}\label{dec}
T_{\mu\nu}= T_{\mu\nu}^{TT}+{1\over 4}\gamma_{\mu\nu}T+{1\over 3}
P_{\mu\nu}{1\over {\cal Q}^{(4)}}T.
\end{eqnarray}  
Noticing that 
\begin{equation}
P_{\mu\nu}(\square_4+8H^2)\varphi=\square_4 P_{\mu\nu}\varphi~,
\end{equation}
for any scalar $\varphi$ 
(which implies that $P_{\mu\nu}f({\cal Q}^{(4)})\varphi=
f({\cal O}^{(4)})P_{\mu\nu}\varphi$ for any smooth function $f$)
the term in the parenthesis in (\ref{delg}) vanishes.
 
Next, let us introduce the Lichnerowicz operator  $\Delta_L$ with
the following properties
\begin{eqnarray}\label{prop}
(\Delta_L -4H^2)T_{\mu\nu}^{TT}&=&{\cal O}^{(4)}T_{\mu\nu}^{TT}~,\nonumber\\
(\Delta_L -4H^2)\gamma_{\mu\nu}\varphi&=&\gamma_{\mu\nu}{\cal Q}^{(4)}
\varphi~,\\
(\Delta_L -4H^2)P_{\mu\nu}\varphi&=&P_{\mu\nu}{\cal Q}^{(4)}\varphi~,\nonumber
\end{eqnarray}
for an arbitrary scalar $\varphi$. From (\ref{delg}) we obtain
\begin{eqnarray}\label{delg2}
{1\over 2}\delta\tilde g_{\mu\nu\pm}&=&
K_\pm(\Delta_L-4H^2)T_{\mu\nu}^{TT}-{A\over 3 H^2 }
P_{\mu\nu}{1\over {\cal Q}^{(4)}}T+
{A\over 12H^2}\gamma_{\mu\nu}T
\nonumber\\
%&=&K_\pm(\Delta_L-4H^2)T_{\mu\nu}
%-{1\over 4}\gamma_{\mu\nu}K({\cal Q}^{(4)})T
%-{1\over 3}P_{\mu\nu}K({\cal Q}^{(4)}){1\over {\cal Q}^{(4)}}T
%-{A\over 3 H^2 }P_{\mu\nu}{1\over {\cal Q}^{(4)}}T+
%{A\over 12H^2}\gamma_{\mu\nu}T
%\nonumber\\
&=&K_\pm(\Delta_L-4H^2)T_{\mu\nu}+{1\over 12}
\gamma_{\mu\nu}\left({\square_4\over {\cal Q}^{(4)}}-3\right)
K_\pm({\cal Q}^{(4)})T-{A\over 3}
\gamma_{\mu\nu} {1\over {\cal Q}^{(4)}}T\nonumber\\&&
-{1\over 3}\nabla_\mu\nabla_\nu\left(
K_\pm({\cal Q}^{(4)})+{A\over H^2}\right){1\over {\cal Q}^{(4)}}T~.
\end{eqnarray}
To get to the second line of (\ref{delg2}) we expressed $T_{\mu\nu}^{TT}$ 
in terms of 
$T_{\mu\nu}$ and $T$ using (\ref{dec}) and the properties (\ref{prop}) of 
$\Delta_L$.
In the last term of the third line, the tachyonic pole ${\cal Q}^{(4)}=0$
could  lead to a boundary divergence, 
impeding integration by parts when we contract with a conserved 
$T'_{\mu\nu}$. 

Below we discuss the two cases of the $(-)$ and 
$(+)$ solutions separately. In the $(-)$ case (this is the 
only case  that has been so far discussed in the literature), 
there is the double pole present in the Green's function. 
Usually  a double pole can be recast into 
a sum/difference of two simple poles, giving rise to ghosts or tachyons. 
In the present case, this decomposition cannot be done 
so clearly. This is because  nonlocal operators
(such a square roots of the covariant d'Alembertian) are involved.
The residue of the double pole diverges,
which may or may not be a signal of a singularity 
of the perturbative approach at scales $\sim H$.
On the other hand, the repulsive nature of the additional 
contributions (as compared to massive gravity) may be suggestive 
of a ghost-like state.  However,  
given the results of Section 2, it 
is not entirely clear whether this ghost-like state, even if present, 
can be emitted in  a final state, or it only appears as an 
intermediate state. In the latter case it would have been 
harmless. To clarify this, 
one needs a BRST invariant construction of the physical Hilbert 
space. Since the truncated theory considered here has 
only a limited interest anyway, we won't pursue this tedious 
project. One is certain, the conclusion reached in the literature 
that the amplitude has a ghost, is not entirely obvious
beyond a reasonable doubt.

For the $(+)$ solution, the $1/{\cal Q}^{(4)}$ 
divergence is not present 
since $K_+({\cal Q}^{(4)})+A/H^2$ also vanishes in the ${\cal Q}^{(4)}\to 0$ 
limit. Therefore, integrating on both sides of (\ref{delg2})
$\int d^4 x \sqrt{\gamma}~T^\prime_{\mu\nu}\cdots$, for the $(+)$ solution,
(and integrating by parts the last term)
\begin{eqnarray}\label{amplit}
{1\over 2}\int  d^4 x \sqrt{\gamma}~T^{\prime\mu\nu}\delta\tilde g_{\mu\nu_+}
&=&\int  d^4 x \sqrt{\gamma}~\Bigg(T^{\prime \mu\nu}K_+
(\Delta_L-4H^2)T_{\mu\nu}
\nonumber\\&&+
{1\over 12}T^{\prime}
\left({\square_4\over {\cal Q}^{(4)}}-3\right)K_+({\cal Q}^{(4)})T
-{A\over 3}
T^\prime {1\over {\cal Q}^{(4)}}T\Bigg)~.
\end{eqnarray}
The pole at ${\cal Q}^{(4)}=0$ has zero residue, and the residue of the 
pole at ${\cal Q}^{(4)}=2H^2$ ($\square_4=-6H^2$) together with the one at 
$\Delta_L =6H^2$ give a {\it massless} graviton amplitude!
Both, the tensorial structure in the numerator and, very importantly, 
the  pole in the denominator, correspond to a propagating
massless spin-2 state on dS background. What is this massless state?
It corresponds to non-normalizable zero-mode discussed in Section 2.
To understand this result better, it is instructive to look at the 
spectral representation of the amplitude (\ref{amplit}).
Let us rewrite (\ref{amplit}) as
\begin{eqnarray}\label{amp}
{\cal A}=\int T^{\prime \mu\nu}K_+\left( \Delta_L-4H^2\right)T_{\mu\nu}-
{1\over 3}T^\prime K_+({\cal Q})T-{1\over 3}T^\prime {H^2\over 
{\cal Q}}K_+({\cal Q})T-{1\over 3}T^\prime{1\over {\cal Q}}T.
\end{eqnarray}
Which can be compared with the amplitude for massive gravity on de Sitter 
space found in \cite{Porrati}:
\begin{eqnarray}\label{amp2}
{\cal A}_M&=&\int T^{\prime \mu\nu}\left( \Delta_L-4H ^2-M^2 -2H^2\right)^{-1}
T_{\mu\nu}-
{1\over 3}T^\prime \left({\cal Q}+M^2 -2 H^2\right) ^{-1}T-\nonumber\\ &&
 {1\over 3}T^
\prime {H^2\over 
{\cal Q}}\left({\cal Q}+M^2-2H ^2\right) ^{-1}T-{H^2\over 6H^2-3M^2 }T^
\prime{1\over {\cal Q}}T.
\end{eqnarray}

{}It is sufficient at this point to obtain the spectral representation at $y=0$
which is completely determined by the function $K_+(z)$ for the first two 
terms in the r.h.s of (\ref{amp}). 
Below, $z$ will refer to ${\cal Q}$,
and we will define the complex coordinate $w=-H^2/4-z$ which 
in the flat space-time limit is positive in the physical domain, 
{\it i.e.}, $w\to-p^2\ge 0$.
Therefore, we study the spectral decomposition of 
$\tilde K_+(w)=K_+(-w-{H^2\over 4})$ in the complex $w$ plane.
\begin{eqnarray}\label{HK}
\tilde K_+(w)={1\over -w-{H^2\over 4}-{Hm_c\over 2}\left(1+{2\over H}\sqrt{-w}
\right)},
\end{eqnarray}
where we have restored $m_c$ to be able to take the flat limit keeping $m_c$ 
finite.
The flat limit of this expression, $\tilde K_+^0(w)$, gives the flat 
propagator of the $(+)$ branch described before:
\begin{eqnarray}\label{fK}
\tilde K_+^0(w)={1\over -w-m_c\sqrt{-w}}.
\end{eqnarray}
This flat propagator can be decomposed as a sum over KK contributions with 
spectral density $\rho$ via a contour integral which goes around the cut just 
below and above the $w$ positive real semi-axis and 
closes at infinity through a
counterclockwise circle (the only contribution coming from the jump of 
the imaginary part of $K$ across the cut, $\Delta {\rm Im} K$):
\begin{eqnarray}
\tilde K_+^0(w)=\int_0^\infty {ds\over s-w} \rho(s)=
{1\over 2\pi}\int_0^\infty
{ds\over s-w} \Delta {\rm Im}K(s)=
{1\over \pi}\int_0^\infty{ds\over s-w}{m_c \sqrt{s}\over s^2+m_c^2 s},
\end{eqnarray} 
where (\ref{fK}) has to be interpreted as 
\begin{eqnarray}
\tilde K_+^0(w)={1\over -w-m_c\sqrt{{\rm e}^{i \pi} w}}
={1\over -w-i m_c\sqrt{w}},
\end{eqnarray}
that implies that there is no pole for $w=|w|{\rm e}^{i \phi}$ in the first 
Riemann sheet, {\it i.e.}, $0\le \phi<2\pi$.
The state corresponding to the pole of the propagator (\ref{fK}) at 
$w={\rm e}^{-i\pi}m_c^2$ is on the second Riemann sheet and does not 
contribute to physical amplitudes. Therefore, by a suitable prescription (which
 amounts to a choice of time direction) we are able to find an amplitude with 
no tachyonic instabilities.

However, in trying to extend this argument to the curved background case, we 
face a problem.  The curved generalization for (\ref{HK}) yields the following 
spectral representation:
\begin{eqnarray}\label{Hrep}
\tilde K_+(w)=-{2\over 4w+H^2}+{1\over\pi }\int_0^\infty {ds 
\over s-w}{16 m_c \sqrt{s}\over 
\left(H(H+2m_c)+4s\right) ^2+4m_c^2 s}.
\end{eqnarray}
As in the flat case, the zero mode from the pole at 
$w={\rm e}^{-i\pi} (H+2m_c)^2 /4$ (corresponding to $z=2H^2$ when $m_c=H$) does
not contribute in (\ref{Hrep}) while the KK's $0\le w_{KK}<\infty$ 
($-H^2/2\le z_{KK}<\infty$) and the isolated mode $w={\rm e}^{i\pi} H^2/4$ do. 
This prescription,  also made the flat amplitude tachyon free, has turned the 
curved amplitude into the one of the $(-)$ branch with a ghostlike double pole 
in the third term of (\ref{amp}).

In order to describe the $(+)$ branch amplitude, with only a 
{\it massless} pole an a branch cut of KK's, we must consider the other 
possible prescription, {\it i.e.}, $\sqrt{-w}\equiv \sqrt{{\rm e}^{-i\pi}w}$, 
the decomposition is then
\begin{eqnarray}\label{Hrep2}
\tilde K_+(w)=-{6\over 4w+(H+2m_c)^2}-{1\over\pi }
\int_0^\infty {ds \over s-w}{16 m_c \sqrt{s}\over 
\left(H(H+2m_c)+4s\right) ^2+4m_c^2 s}~.
\end{eqnarray}
Now, the zero mode from the pole of (\ref{HK}) at 
$w={\rm e}^{i\pi} (H+2m_c)^2 /4$ is on the first Riemann sheet and does 
contribute to the decomposition and the continuum enters with opposite sign.
This relative sign implies that once the KK modes are treated as 
conventional positive-residue states, the residue of the zero mode has to be 
negative. This would imply that 
the zero-mode is a ghost. However, the zero-mode is not a normalizable 
state and because of this its interpretation is obscure. In particular,
whether or not it can be emitted in a final state 
by any mater source localized  on the brane is not clear.

\section{Outlook}

We studied the question of small perturbations on the 
self-accelerated solution of the DGP model. The issue is 
rather involved. We showed that the small perturbations on 
an {empty} self-accelerated background can be quantized 
without yielding  ghosts or other instabilities. 
For conformal sources, such as radiation,
small perturbations are instability free as well. 
More realistic, non-conformal sources, however, 
could trigger ghost-like instabilities in the 
linearized theory, and become unstable. We have suggested,  in Section 4,
possible loopholes in the latter conclusion,
even within the linearized theory itself. More importantly,
however, those non-conformal sources lead also
to the breakdown of the linearized calculations, 
and, therefore, no conclusion on the (in)stability 
of the solution can be drawn without invoking 
non-linear dynamics.  For the available non-linear 
spherically symmetric solution  the dangerous linearized 
modes are spurious.

To establish stability of generic non-conformal sources, one has 
to perform calculations of 5D perturbations about a background 
produced by the source itself, since the  latter is dominating 
over the cosmological background locally. One should 
see which of the obtained perturbations could be matched, at the 
scale $r_*$, to solutions in the far-away region. 
Further interesting question 
is to study deviations from the self-accelerated background by 
introducing a small cosmological constant to explore the spectrum 
of the theory. We hope to report on some of these issues elsewhere.

\vspace{0.1in}

\subsection*{Acknowledgments}

We would like to thank G. Dvali, G. Esposito-Far\'ese, T. Gherghetta, 
M. Kleban, N. Kaloper, K. Koyama, P. Mannheim, J. Mourad, M. Porrati, 
M. Redi, R. Scoccimarro, T. Tanaka, and M. Zaldarriaga for 
useful discussions. Work of GG and AI is 
supported in part by NASA grant NNGG05GH34G and 
in part by NSF grant 0403005. CD would like to thank CCPP at 
New York University, and GG to thank the Weizmann Institute, for 
hospitality, where parts of this work were done.

%\newpage

\section*{Appendix A}

The goal of the present section is to discuss a toy model of a 
real  scalar field minimally coupled to 
a gravitational self-accelerated background.
The idea is that the spectrum of small perturbations of 
this scalar bears some similarities to that of the 
gravitational perturbations.  The action of the toy model is 
\beq
S= \int d^4x dy \sqrt{g} \left (-{1\over 2} g^{AB} 
\partial _A \Phi \partial_B \Phi 
\right ) + 2 r_c \int d^4x \sqrt {\gamma}  
\left (-{1\over 2} \gamma^{\mu\nu} 
\partial _\mu \varphi \partial_\nu \varphi 
\right )\,,
\label{saction}
\eeq
where $A,B =0,1,2,3,{\it 5}$, $\mu,\nu =0,1,2,3,$, 
$\varphi(x) \equiv \Phi (x, y=0)$, 
$\gamma_{\mu\nu}(x) = g_{\mu\nu}(x,y=0)$, 
and the metric of the spatially-flat self-accelerated background 
is \cite {Cedric} 
\beq
ds^2 = (1+H|y|)^2 \left (-dt^2 + a^2(t)d{\vec x}^2 \right )+dy^2 \equiv 
A^2(y) \gamma_{\mu\nu}(x) dx^\mu dx^\nu + dy^2\,.
\label{bcground}
\eeq
The equation of motion reads:
\beq
\partial_A\left ( \sqrt{g} g^{AB} \partial_B \Phi \right)
+2 r_c \delta (y) \partial_\mu\left ( \sqrt{\gamma} \gamma^{\mu\nu} 
\partial_\nu \varphi \right) =0\,.
\label{seq}
\eeq
Since $\sqrt{g} =A^4a^3$ and $\sqrt {\gamma} = a^3$ we obtain
for the bulk equation:
\beq
{1\over \sqrt{\gamma}} \partial_\mu \left ( \sqrt{\gamma} \gamma^{\mu\nu}
\partial_\nu \Phi \right ) \equiv \square_4 \Phi  = 
-{\partial_y (A^4 \partial_y \Phi)\over A^2}\,.
\label{sbulk}
\eeq
Moreover, for the junctions condition we get
\beq
[\partial_y \Phi]^{+\epsilon}_{-\epsilon} = - 2 r_c \square_4 \Phi |_0\,.
\label{sjunction}
\eeq
We look for the KK modes that are solutions to (\ref {sbulk}) and  
(\ref {sjunction}) and can be expressed in the form:
\beq
\Phi (x,y) \equiv \int \varphi_m(x) f_m(y) dm\,,~~~~~\square_4 \varphi_m(x)
=m^2 \varphi_m(x)\,.
\label{sKK}
\eeq
For the KK modes the bulk equation (\ref {sbulk}) and the  junction 
condition (\ref {sjunction}) can be rewritten as follows:
\beq
m^2 f_m(y) = -{\partial_y (A^4 \partial_y f_m)\over A^2}\,,
\label{sbulkKK}
\eeq
and 
\beq
[\partial_y f_m(y) ]^{+\epsilon}_{-\epsilon} = - 2 r_c m^2 f_m(0)\,.
\label{sjunctionKK}
\eeq
As in the graviton case, and because of the boundary condition 
that depends on the  mass of a KK mode (\ref {sjunctionKK}), 
the modes themselves are not orthogonal. 

To solve the above equations it is useful to make the following 
change of variables from $y$ to $z$, and introduce a new functions
$u_m$:
\beq
dz \equiv {dy \over A(y)}\,, ~~~~ f_m \equiv A^{-3/2} u_m \,. 
\label{snew}
\eeq
In terms of the new coordinate $ A(z) = {\rm exp} ( H|z|)$,
and $\partial_y = A^{-1}(z) \partial_z$. As a result, the bulk equation
takes the form:
\beq
-{d^2u_m  \over dz^2}  + \left ({9H^2\over 4} -m^2  \right )u_m =0\,.
\label{sbulku}
\eeq
The junctions condition (\ref {sjunctionKK}) 
rewritten in terms of $u_m$ reads:
\beq
[\partial_z u_m(z)]^{+\epsilon}_{-\epsilon} = \left 
( 3H - 2 r_c m^2 \right )u_m(0)\,.
\label{sjunctionu}
\eeq
From the above two equations it is immediately clear that there is a 
zero-mode solution $m=0$, $u={\rm exp}(3H|z|/2)$. However, 
this mode is diverging in the bulk. 

Other solutions of the above system of  equations should be supplemented 
by the normalization  condition, which in terms of the old and new variables 
reads as follows:
\beq
\int^{+\infty}_{-\infty}  dy {\sqrt{g}\over \sqrt{\gamma}} g^{00} f_m^2(y)
= \int^{+\infty}_{-\infty}  dz\, u_m^2(z) < \infty \,. 
\label{snorm}
\eeq

Let us analyze the above system of equations. 
It is straightforward to check that there exists only one 
normalizable solution:
\beq
u_{m_*} = \sqrt{\alpha}\, e^{ - \alpha |z| }\,,~~~~~
\alpha\equiv {3H\over 2} -m_c\,,
\label{ssol}
\eeq
while the mass of this mode is
\beq
m_*^2 = 3Hm_c -m_c^2 \,.
\label{smass}
\eeq
The above corresponds to the 
self-accelerated background if $H=m_c$. 
Then, we get $\alpha = H/2$, $m_*^2 = 2H^2$, and   
$u_{m_*}= ({H/ 2 A})^{1/2}$.

What are the other solutions of (\ref {sbulku}) and  (\ref {sjunctionu})?
The rest of the solutions constitute a 
continuum of massive states with $m^2 \ge 9H^2 /4$ with the 
plane-wave normalizable wavefunctions:
\beq
u_m(z) = A sin (\omega |z|) + B cos (\omega z)\,, ~~~ \omega 
\equiv \sqrt {m^2 -{9H^2\over 4}}\,,
\label{splane}
\eeq 
where the normalization coefficients $A$ and  $B$ 
can be calculated
\beq
A = - B\, { 2 m^2 -3m_c H  \over m_c \omega}\,,
\label{sAvsB}
\eeq 
while the expression for $B$  determines an important  dependence  
of the wavefunctions on $m^2$ on the brane, in analogy with the flat case 
\cite {Nitti}.

\section*{Appendix B}

In this appendix we consider linearized 
4D massive gravity about a dS space with the special relation
between the mass and dS curvature given in (\ref {mR}).
The Lagrangian density takes the form 
(see, e.g., \cite {Higuchi}, but the overall coefficient 
is different here)
\beq
-{1\over 2} {\cal L}_* = {1\over 2} (\nabla_\mu h^{\mu\alpha})^2  +
{1\over 4}h_{\mu\nu} \square_4  h^{\mu\nu} 
-{1\over 4} h \square_4 h +{1\over 2} h^{\mu\nu} \nabla_\mu \nabla_\nu h 
\nonumber \\
- {1\over 2} H^2 \left ( h^2_{\mu\nu} +{1\over 2} h^2  \right ) 
- {1\over 4} m_*^2 \left (   
h^2_{\mu\nu} -h^2 \right ) \,.
\label{4DmassdS}
\eeq
For a generic value of the graviton mass the above Lagrangian 
has no gauge symmetry.  However, for a special  
value (\ref {mR}) there is a  symmetry present:
\beq
\delta h_{\mu\nu}(x) = 
(\nabla_\mu \nabla_\nu + H^2 \gamma_{\mu\nu}) \rho (x)\,.
\label{enhanced}
\eeq
The above gauge transformation  allows one to gauge-remove 
one out of five on-shell polarizations of the 
massive graviton \cite {Deser}. However, this theory does not admit
a coupling to non-conformal matter.
There are many ways to see this. The most straightforward one 
is to write down the 
equations of motion, use the Bianchi identities, and observe that 
the trace equation would necessarily imply $T^\mu_\mu=0$. This  
can also be established by noticing that introduction of the 
coupling $h_{\mu\nu}T^{\mu\nu}$ in the Lagrangian (\ref {4DmassdS}) 
violates the symmetry (\ref {enhanced}) unless $T^\mu_\mu=0$.
Then, one could introduce a coupling {\it a la} St\"uckelberg
$(h_{\mu\nu}- (\nabla_\mu \nabla_\nu + H^2 \gamma_{\mu\nu})\sigma) 
T^{\mu\nu}$, where $\sigma$ is a field that also transforms 
as $ \delta \sigma(x) = \rho(x)$ when the metric is transformed according to 
(\ref {enhanced}). Such a coupling would not violate the symmetry of 
the theory. However, the equation of motion of the $\sigma$ field would 
require that $T^\mu_\mu=0$. As we have seen in the 
Section 2, in the DGP model this difficulty is avoided
due to the extrinsic curvature terms.

To this end we would like to calculate the propagator 
of the theory given by (\ref {4DmassdS}) without coupling it to any matter. 
The gauge independent part of the propagator would 
tell us how many degrees of freedom are propagating in the theory.
Without introducing the source, the propagator is defined
as an inverse of the quadratic operator in (\ref {4DmassdS}).
However, because of the symmetry  (\ref {enhanced}), this operator is 
not invertible. Hence, we should 
proceed by introducing the gauge fixing term which we choose to be 
\beq
{\cal L}_{\rm gf} = - {1\over \alpha} h^2\,.
\label{gf}
\eeq
After this term is included, 
we introduce a ``spectator'' $T_{\mu\nu}$,
for which $T\neq 0$. Formally, this is possible because of the gauge fixing 
term. However we should keep in mind that the original theory
wouldn't  admit any non-conformal matter,  and, in that respect,
the  introduction of the ``spectator'' non-conformal 
source  after the gauge fixing is just a technical trick that  
enables one to extract the non-derivative terms of the 
propagator without too much of calculations. We follow this philosophy 
below.

In this gauge the equations of motion read:
\beq
{1\over 2} \left (\square_4 h_{\mu\nu}  - 
\nabla_\nu \nabla_\alpha h^\alpha_\mu -\nabla_\mu \nabla_\alpha h^\alpha_\nu
+ \nabla_\mu \nabla_\nu h  \right )+ {1\over 2}\gamma_{\mu\nu} 
\left ( \nabla_\alpha \nabla_\beta h^{\alpha\beta} -\square_4 h \right ) 
 \nonumber \\
- H^2 \left (h_{\mu\nu} +{1\over 2} \gamma_{\mu\nu} h \right ) - {m_*^2\over 2}
\left ( h_{\mu\nu} - \gamma_{\mu\nu} h \right ) +
{1\over \alpha }\gamma_{\mu\nu}  h = - T_{\mu\nu}\,.
\label{masseq}
\eeq
The Bianchi identities enforce the following constraint:
\beq
{m_*^2\over 2} ( \nabla^\mu h_{\mu\nu} -\nabla_\nu h) = - 
{1\over \alpha }\nabla_\nu h \,.
\label{Bianchi}
\eeq
The trace of equation (\ref {masseq})  takes the form
\beq
\nabla^\mu \nabla^\nu h_{\mu\nu} -\square_4 h + 3\left ({m_*^2\over 2} -H^2   
\right ) h + {4\over \alpha }h  = - T\,. 
\label{trace}
\eeq
The above system of equations can be solved. In what follows 
we concentrate on the case when (\ref {mR}) is satisfied, i.e.,
$m_*^2=2H^2$ . First we solve for the trace 
\beq
h = {\alpha TH^2\over \square_4 - 4H^2}.
\label{h}
\eeq
As we see, in a theory without the gauge fixing term 
(i.e, when $\alpha \to \infty$) the field diverges.

Next using the identity (\ref {master}) we solve for $h_{\mu\nu}$: 
\beq
-{1\over 2} {h}_{\mu\nu} = {T_{\mu\nu} - {1\over 2}\gamma_{\mu\nu} T \over
\square_4- 4H^2 } + \left (1 - {\alpha H^2\over 2}\right )
{1\over (\square_4 - 4H^2)} ( \nabla_\mu \nabla_\nu - 
H^2 \gamma_{\mu\nu}) {T\over
(\square_4 - 4H^2)}\,. 
\label{hbarresult}
\eeq
Fortunately, the double pole terms on the r.h.s. 
are gauge dependent\footnote{Notice that the gauge-dependent 
terms in this case contain not only derivatives but also terms 
proportional to $\gamma_{\mu\nu}$. This is a direct 
consequence of the form of the gauge transformations 
(\ref {enhanced}).}. On the other hand, the gauge-independent part of 
the above expression gives  the non-derivative part of the 
inverse of the gauge kinetic term of (\ref {4DmassdS}). The latter 
takes the form
\beq
{1\over 2} {\gamma_{\mu\alpha} \gamma_{\nu\beta} + 
\gamma_{\mu\beta} \gamma_{\nu\alpha}
- \gamma_{\mu\nu} \gamma_{\alpha\beta} \over
\square_4- 4H^2 }+...
\label{propmR}
\eeq
The tensorial structure in the numerator of Eq. (\ref {propmR}) 
confirms that the theory (\ref {4DmassdS})
at the linearized level does not propagate the fifth,
helicity-0, degree of freedom \cite {Deser}.

\section*{Appendix C}

In this appendix we give the expressions used in Section 3 for the 
first and second order perturbations of the Ricci tensor around a de 
Sitter background metric $\gamma_{\mu\nu}$. 

The metric is expanded as    
\beq
g_{\mu\nu}=\gamma_{\mu\nu}+\delta g_{\mu\nu}~,
\eeq
with inverse
\beq
g^{\mu\nu}=\gamma^{\mu\nu}-\delta g^{\mu\nu}+\delta g^\mu_\rho\delta 
g^{\rho\nu}\,.
\eeq
The first order Ricci tensor components are given by
\beq\label{r1}
R_{\mu\nu}^{(1)}(\delta g)=-{1\over 2}\left(\Box \delta g_{\mu\nu}-
\nabla_\alpha\nabla_\mu\delta g^\alpha_\nu-
\nabla_\alpha\nabla_\nu\delta g^\alpha_\mu+
\nabla_\mu\nabla_\nu\delta g^\alpha_\alpha\right)~,
\eeq
which for a transverse traceless perturbation, 
$\delta g_{\mu\nu}=h_{\mu\nu}^{TT}$, reduces to
\beq\label{r1tt}
R_{\mu\nu}^{(1)}(h^{TT})=-{1\over 2}\Box h^{TT}_{\mu\nu}+4H^2h_{\mu\nu}^{TT}~.
\eeq
Note that the $\nabla_\alpha$ has to be commuted through $\nabla_{\mu, \nu}$
in the right hand side of (\ref{r1}) to get (\ref{r1tt}). Here, we
recall that for the de Sitter background metric $\gamma_{\mu\nu}$ we have
$[\nabla_\alpha, \nabla_\mu]h^\alpha_\nu=
4H^2 h_{\mu\nu}-H^2\gamma_{\mu\nu}h^\alpha_\alpha$. 

The second order terms are given by
\beq\label{r2}
R^{(2)}_{\mu\nu}(\delta g)&=&+{1\over 4} \left(
\nabla^\rho \delta g^\alpha_\alpha-2\nabla^\alpha \delta g^\rho_\alpha
\right)\left(
\nabla_\mu\delta 
g_{\nu\rho}+\nabla_\nu\delta g_{\mu\rho}-\nabla_\rho\delta g_{\mu\nu}
\right)\nonumber\\&&
-{1\over 4} \left(
\nabla^\rho \delta g^\alpha_\mu-\nabla_\mu\delta 
g^{\rho\alpha}-\nabla_\alpha\delta g^\rho_\mu
\right)\left(
\nabla_\alpha\delta g_{\nu\rho}+\nabla_\nu\delta 
g_{\alpha\rho}-\nabla_\rho\delta g_{\nu\alpha}
\right)\nonumber\\&&
-{1\over 2}\delta g^{\alpha\rho}\left(\nabla_\alpha\nabla_\mu\delta 
g_{\rho\nu}+
\nabla_\alpha\nabla_\nu\delta g_{\rho\mu}-\nabla_\alpha\nabla_\rho\delta 
g_{\mu\nu}
\right)\nonumber\\&&
+{1\over 2}\delta g^{\alpha\rho}\left(\nabla_\mu\nabla_\alpha\delta 
g_{\rho\nu}+
\nabla_\mu\nabla_\nu\delta g_{\rho\alpha}-\nabla_\mu\nabla_\rho\delta 
g_{\nu\alpha}
\right)~.
\eeq
Which for a transverse traceless perturbation reads 
(we drop the $TT$ superscript on the r.h.s.)
\beq\label{r2tt}
R^{(2)}_{\mu\nu}(h^{TT})=
-{1\over 4} \left(
\nabla^\rho h^\alpha_\mu-\nabla_\mu h^{\rho\alpha}-\nabla_\alpha h^\rho_\mu
\right)\left(
\nabla_\alpha h_{\nu\rho}+\nabla_\nu h_{\alpha\rho}-\nabla_\rho h_{\nu\alpha}
\right)\nonumber\\ 
-{1\over 2}h^{\alpha\rho}\left(\nabla_\alpha\nabla_\mu h_{\rho\nu}+
\nabla_\alpha\nabla_\nu h_{\rho\mu}-\nabla_\alpha\nabla_\rho h_{\mu\nu}
-\nabla_\mu\nabla_\alpha h_{\rho\nu}-
\nabla_\mu\nabla_\nu h_{\rho\alpha}+\nabla_\mu\nabla_\rho h_{\nu\alpha}
\right)~.
\eeq

\end{document}